\documentclass{ws-wsarpp}
\usepackage{hyperref}
\usepackage[super,sort&compress,comma]{natbib}
\usepackage{slashed}
\usepackage{color}
\usepackage{todonotes}
\usepackage{comment}

\begin{document}

\markboth{L.~Di Luzio, H.~Gisbert, G.~Levati, P.~Paradisi, P.~S{\o}rensen}{CP-Violating Axions: A Theory Review}

%
%
\def\eq#1{{Eq.~(\ref{#1})}}
\def\eqs#1#2{{Eqs.~(\ref{#1})--(\ref{#2})}}
\def\fig#1{{Fig.~\ref{#1}}}
\def\figs#1#2{{Figs.~\ref{#1}--\ref{#2}}}
\def\Table#1{{Table~\ref{#1}}}
\def\Tables#1#2{{Tables~\ref{#1}--\ref{#2}}}
\def\sect#1{{Sect.~\ref{#1}}}
\def\sects#1#2{{Sects.~\ref{#1}--\ref{#2}}}
\def\app#1{{App.~\ref{#1}}}
\def\apps#1#2{{Apps.~\ref{#1}--\ref{#2}}}
\def\e{e}
\def\vev#1{\left\langle #1\right\rangle}
\def\abs#1{| #1|}
\def\mod#1{\abs{#1}}
\def\Im{\mbox{Im}\,}
\def\Re{\mbox{Re}\,}
\def\trace{\mbox{Tr}\,}
\def\Tr{\mbox{Tr}\,}
\def\tr{\mbox{Tr}\,}
\def\det{\mbox{det}\,}
\def\etal{\hbox{\it et al.}}
\def\ie{\hbox{\it i.e.}{}}
\def\eg{\hbox{\it e.g.}{}}
\def\etc{\hbox{\it etc}{}}
\def\diag{\mbox{diag}\,}
\newcommand{\neff}{n_{\rm eff}}
\renewcommand{\d}{{\rm d}}
\renewcommand{\bar}{\overline}
\newcommand{\hoppet}{\textsc{Hoppet}}
\newcommand{\R}[2]{$R_{#1#2}$}
\newcommand{\as}{$\alpha_s$}
\newcommand{\rr}[1]{{\color{red}#1}}
\newcommand{\bb}[1]{{\color{blue}#1}}

\newcommand{\beq}{\begin{equation}}
\newcommand{\eeq}{\end{equation}}
\newcommand{\bea}{\begin{eqnarray}}
\newcommand{\eea}{\end{eqnarray}}

\renewcommand{\[}{\left[}
\renewcommand{\]}{\right]}
\renewcommand{\(}{\left(}
\renewcommand{\)}{\right)}

\newcommand{\LDL}[1]{{\color{red}{\bf #1}}}
\newcommand{\GL}[1]{{\color{blue}{\bf #1}}}

\newcommand{\X}{{\cal X}}
\newcommand{\N}{{\cal N}}
\newcommand{\Q}{{\cal Q}}
\newcommand{\E}{{\cal E}}
\newcommand{\PQ}{{\rm PQ}}

\title{\large CP-Violating Axions: A Theory Review}

\author{Luca Di Luzio$^a$, Hector Gisbert$^{b,a}$, Gabriele Levati$^{b,a}$, Paride Paradisi$^{b,a}$, Philip S{\o}rensen$^{b,a}$}

\address{$^a$Istituto Nazionale di Fisica Nucleare (INFN), Sezione di Padova, \\
Via F. Marzolo 8, 35131 Padova, Italy \\
$^{b}$Dipartimento di Fisica e Astronomia `G.~Galilei', Universit\`a di Padova,
 \\ Via F. Marzolo 8, 35131 Padova, Italy
}

\maketitle


\begin{abstract}
We review the physics case for CP-violating axions. 
In the first part, we focus on the Quantum Chromodynamics (QCD) axion and argue that new sources of CP violation beyond QCD misalign the axion solution to the strong CP problem and can manifest themselves via a tiny scalar axion-nucleon component. We hence highlight recent advancements in calculating this scalar axion-nucleon coupling, a parameter that could be probed via 
axion-mediated force experiments. 
In the second part, we focus on axion-like particle (ALP) interactions entailing the most general sources of CP violation. After classifying the full set of CP-violating Jarlskog invariants, we report on recent calculations of ALP contributions to permanent electric dipole moments. 
We finally speculate on possible ultraviolet completions of the CP-violating ALP.
\end{abstract}



\clearpage

\section{Introduction}
\label{sec:intro}
Among light and weakly coupled new physics 
scenarios beyond the Standard Model (SM),
the Quantum Chromodynamics (QCD) axion stands out as a most compelling paradigm, by simultaneously providing an elegant solution to the strong CP 
problem via the Peccei Quinn (PQ) mechanism
\cite{Peccei:1977hh,Peccei:1977ur,Weinberg:1977ma,Wilczek:1977pj}
and an excellent dark matter candidate \cite{Preskill:1982cy,Abbott:1982af,Dine:1982ah}. 
The experimental landscape for axion searches 
has considerably broadened 
in the recent years, see \cite{Irastorza:2018dyq,Sikivie:2020zpn} for recent reviews. 
While traditional axion searches target the standard
axion-photon coupling, 
novel detection strategies have been also proposed to 
probe axion couplings to nucleons \cite{Graham:2013gfa,Budker:2013hfa,Arvanitaki:2014dfa,Geraci:2017bmq,DiLuzio:2021qct} and electrons \cite{Barbieri:2016vwg,Crescini:2016lwj,Crescini:2017uxs,QUAX:2020adt}.
These are especially relevant in view of their complementarity in disentangling different realizations of QCD axion models.
Other low-energy signatures of QCD axions could stem from flavour- 
and CP-violating processes. Indeed, although negligible in benchmark axion models \cite{Kim:1979if,Shifman:1979if,Zhitnitsky:1980tq,Dine:1981rt}, 
sizeable flavour- and CP-violating axion couplings may be naturally 
generated in non-minimal scenarios addressing the SM flavour puzzle~\cite{Davidson:1981zd,Wilczek:1982rv,Berezhiani:1983hm,Ema:2016ops,Calibbi:2016hwq,Bjorkeroth:2018ipq} 
and/or accounting for the observed matter-antimatter asymmetry of the universe.

Standard QCD axion models predict a strict relation between the axion mass and its decay constant, enforcing an upper bound on the axion mass of about 0.1 eV after imposing astrophysical and cosmological bounds. Instead, in the more general framework of axion-like particles (ALPs), such a stringent condition is relaxed and
much larger ALP masses can be conceived.
ALP interactions with SM fermions and gauge bosons are then described 
via an effective Lagrangian containing operators up to 
$d=5$ \cite{Georgi:1986df}. 
The opportunity to look for ALPs with masses above the MeV scale, whose couplings are not tightly constrained by astrophysical bounds, opens the possibility to probe ALP interactions via collider searches~\cite{Mimasu:2014nea,Jaeckel:2015jla,Knapen:2016moh,Brivio:2017ije,Bauer:2017nlg,Bauer:2017ris,Mariotti:2017vtv,Bauer:2018uxu,Aloni:2018vki,Merlo:2019anv} and flavour-violating observables, both in the quark~\cite{Batell:2009jf,Gavela:2019wzg,MartinCamalich:2020dfe,Guerrera:2021yss,Bauer:2021mvw,DiLuzio:2023ndz,Cornella:2023kjq}
and in the lepton sectors~\cite{Bauer:2019gfk,Cornella:2019uxs,Calibbi:2020jvd}. 

Conversely, CP-violating observables involving 
either QCD axions or ALPs
have received relatively less attention. 
It is hence the purpose of this review to summarize the physics case for CP-violating axions. In the case of QCD axions, this is addressed by providing the basic ingredients behind the origin of CP-violating axion couplings, and by reporting on recent calculations of CP-violating couplings to nucleons. 
Instead, in the case of ALPs, we discuss the most general CP-violating ALP interactions, both for ALP masses above few GeV, where QCD can be treated perturbatively, and below the GeV scale, where non-perturbative methods such as chiral perturbation theory ($\chi$pt) are mandatory. 
The impact on electric dipole moments (EDMs) is discussed in both regimes.

The experimental prospects for probing CP-violating axion interactions via axion-mediated force experiments and future improvements in the measurements of permanent EDMs of nucleons, nuclei, atoms and molecules
will be also reviewed.
The review is structured around two main parts, discussing 
respectively the case of CP-violating QCD axions
(\sect{sec:QCDaxions}) and CP-violating ALPs (\sect{sec:ALPs}).

\section{CP-violating QCD axions}
\label{sec:QCDaxions}

The QCD axion originally emerged from the need to 
``wash out'' CP violation from strong interactions \cite{Peccei:1977hh,Peccei:1977ur,Weinberg:1977ma,Wilczek:1977pj}. 
From an effective field theory (EFT) perspective, 
the PQ solution to the strong CP problem can be formulated as follows. 
The SM Lagrangian is augmented by a spin-0 field $a(x)$ endowed  with a pseudo-shift symmetry, 
$a \to a + \alpha f_a$, 
that is broken only by the operator 
\beq 
\label{eq:aGGtilde}
\frac{a}{f_a} \frac{\alpha_s}{8\pi}  G \tilde G 
~ , 
\eeq
with $G\tilde G \equiv \frac{1}{2} \epsilon^{\mu\nu\rho\sigma} G^a_{\mu\nu} G^a_{\rho\sigma}$. 
After employing the pseudo-shift 
symmetry to reabsorb the QCD $\theta$ term by setting $\alpha = - \theta$, one is left 
in the shifted theory
with the  
operator in \eq{eq:aGGtilde}. Hence, the question of CP violation in strong interactions 
is traded for 
a dynamical question about the vacuum expectation value (VEV) of the axion field
\beq 
\theta_{\rm eff} \equiv \frac{\vev{a}}{f_a} ~, 
\eeq
resulting
into an effective $\theta$ parameter,  
with 
$\abs{\theta_{\rm eff}} \lesssim 10^{-10}$ 
from the non-observation of the 
neutron EDM (nEDM)~\cite{Abel:2020pzs,Pendlebury:2015lrz}. 

A general result, due to Vafa and Witten \cite{Vafa:1984xg}, 
ensures that the ground state energy density 
of QCD $\E$ is minimized for $\theta_{\rm eff} = 0$,  
namely $\E(0) \leq \E (\theta_{\rm eff})$. 
The argument is based on an inequality that exploits the 
path-integral representation of $\E$ \cite{Coleman:1985rnk}. 
At large 4-volume in Euclidean space,
one has 
\begin{align}
\label{eq:VF}
e^{-V_4 \E (\theta_{\rm eff})} &= 
\int \mathcal{D} \varphi e^{-S_0 + i \theta_{\rm eff} \Q}  
= \Big| \int \mathcal{D} \varphi e^{-S_0 + i \theta_{\rm eff} \Q} \Big| \nonumber \\
&\leq 
\int \mathcal{D} \varphi \Big| e^{-S_0 + i \theta_{\rm eff} \Q} \Big| 
= e^{-V_4 \E (0)} ~ ,
\end{align}
where $\varphi$ denotes the collection of QCD fields (quarks and gluons), 
$S_0$ is the QCD Lagrangian in absence of the $\theta$ term 
and we introduced the topological charge operator $\Q \equiv \int d^4x \frac{\alpha_s}{8\pi}  G \tilde G$, 
which takes integer values in the background of QCD instantons. 
A crucial assumption, on which the proof in \eq{eq:VF} relies,
consists in the positive definiteness of the path-integral measure 
\beq 
\mathcal{D} \varphi \equiv \mathcal{D} A^a_\mu ~ \det(\slashed{D} + M) ~ , 
\eeq
where the Gaussian path-integral over the fermionic fields has been explicitly performed. 
However, while the fermionic determinant is positive definite in a vector-like theory such as  
QCD,\footnote{Given a non-zero eigenvalue $\lambda$ 
of the Euclidean Dirac operator $i\slashed{D} = i \sum_{\alpha=1}^4\gamma^\alpha D_\alpha$, 
that is $i\slashed{D} \psi = \lambda \psi$, 
then also $-\lambda$ is an eigenvalue, 
since $i\slashed{D} (\gamma_5 \psi) = - \lambda (\gamma_5 \psi)$. 
Hence, starting from the Euclidean Dirac Lagrangian $\bar\psi (\slashed{D} + M) \psi$,
the fermionic determinant 
reads 
$\det(\slashed{D} + M) = \Pi_{\lambda} (M - i \lambda)
= \Pi_{\lambda > 0} (M - i \lambda) (M + i \lambda) 
= \Pi_{\lambda > 0} (M^2 + \lambda^2) > 0$ 
(see e.g.~\cite{Vafa:1983tf}).} 
this is not the case for a chiral theory like the SM.   
Hence, we cannot apply the Vafa-Witten theorem to the 
SM, although the argument does not automatically imply that 
$\theta_{\rm eff} \neq 0$ in the SM. An extra ingredient of 
the SM is that CP is explicitly broken in the quark sector by 
the Cabibbo-Kobayashi-Maskawa (CKM) phase, which is responsible 
for a $\theta_{\rm eff} \neq 0$ term.

\subsection{The QCD axion ground state in the presence of CP violation}
  
Given a short-distance CP-violating local operator $\mathcal{O}_{\rm CPV} (x)$, 
we can estimate the value of $\theta_{\rm eff}$  
by expanding the axion potential as 
\beq 
\label{eq:Vthetaeff}
V(\theta_{\rm eff}) = K' \theta_{\rm eff} + \frac{1}{2} K \theta_{\rm eff}^2 + \mathcal{O} (\theta^3_{\rm eff}) ~ , 
\eeq 
and focus on the $\theta_{\rm eff} \ll 1 $ regime, since we know 
that $\abs{\theta_{\rm eff}} \lesssim 10^{-10}$. 
Here, $K$ is a 2-point function also known as 
topological susceptibility \cite{Shifman:1979if}
\beq
\label{eq:defK}
K = i \int d^4 x 
\langle 0 | \mathcal{T} \frac{\alpha_s}{8\pi} G\tilde G (x) \frac{\alpha_s}{8\pi} G\tilde G (0) | 0 \rangle 
~ , 
\eeq
where $\mathcal{T}$ denotes the time-ordered product. 
This quantity is related to the axion mass via $K = m_a^2 f_a^2$,
while $K'$ is a 1-point function given by \cite{Pospelov:1997uv}
\beq
\label{eq:defKp}
K' = i \int d^4 x 
\langle 0 | \mathcal{T} \frac{\alpha_s}{8\pi} G\tilde G (x) \mathcal{O}_{{\rm CPV}} (0) | 0 \rangle 
~ . 
\eeq
Note that $K' \neq 0$ because $G\tilde G$ and $\mathcal{O}_{{\rm CPV}}$ are both CP-odd, 
and CP-violating effects can be safely neglected for the evaluation of the QCD matrix element. 
The induced $\theta_{\rm eff}$ is hence obtained via 
the direct minimization of $V(\theta_{\rm eff})$ in \eq{eq:Vthetaeff}, yielding 
\beq
\label{eq:thetaeffmin}
\theta_{\rm eff} \simeq - \frac{K'}{K} ~ .
\eeq
The value of $\theta_{\rm eff}$ in the SM was estimated by Georgi and Randall in Ref.~\cite{Georgi:1986kr}. 
At energies below $\Lambda_{\chi} = 4\pi f_\pi$ with $f_\pi \simeq 92$ MeV one can write a flavour conserving, 
CP-violating operator
\beq 
\mathcal{O}^{\rm SM}_{\rm CPV} = \frac{G_F^2}{m_c^2} J_{\rm CKM} 
[\bar u \gamma^\mu (1-\gamma_5) d \cdot \bar d \gamma_\mu] \slashed{D} 
[\gamma^\nu (1-\gamma_5) s \cdot\bar s \gamma_\nu (1-\gamma_5) u] ~ ,
\eeq
which is obtained in the SM after integrating out the $W$ boson and the charm quark 
(see diagram in Fig.~1 of \cite{Georgi:1986kr}),  
while $J_{\rm CKM} = \Im V_{ud}V^*_{cd}V_{cs}V^*_{us} \simeq 3 \times 10^{-5}$ is the reduced Jarlskog invariant \cite{Jarlskog:1985ht}. 
A proper evaluation of the $K'$ matrix element in the presence of $\mathcal{O}^{\rm SM}_{\rm CPV}$ is a non-trivial task.  
However, by the rules of naive dimensional analysis (NDA) \cite{Manohar:1983md}
one expects $K' \sim f_\pi^{4} \frac{G_F^2}{m_c^2} J_{\rm CKM} f_\pi^{4} \Lambda^2_\chi$ 
\cite{Georgi:1986kr}.\footnote{The extension of NDA power counting 
arguments to CP-violating operators beyond the SM can be found in Ref.~\cite{Barbieri:1996vt}.} 
Hence, taking 
also 
$K \sim f_\pi^4$ (indeed,  
$K \simeq (76 \ \text{MeV})^4$ -- see  \cite{diCortona:2015ldu,Borsanyi:2016ksw}),  
one obtains 
\beq
\label{eq:thetaeffmin2}
\theta^{\rm SM}_{\rm eff} \sim \frac{G_F^2}{m_c^2} J_{\rm CKM} f_\pi^4 \Lambda^{2}_{\chi} \sim 10^{-19} ~ ,  
\eeq
which should be taken as an indicative estimate 
of the CKM contribution.  

It is interesting to note that 
the PQ mechanism would have been ineffective in a modified SM 
with 
the QCD and the Fermi scales 
closer to each other 
and/or in the presence of a trivial flavour structure such that $J_{\rm CKM} \sim 1$. 
Also, the PQ mechanism is not generically going to work in a low-scale theory beyond the SM 
with generic CP violation. To see this, assume a $d=6$ CP-violating operator 
$\mathcal{O}_{\rm CPV}/\Lambda_{\rm BSM}^2$, 
coupled to QCD.  
Following an estimate similar to that in \eq{eq:thetaeffmin2} one obtains 
\beq 
\label{eq:thetaeffmin3}
\theta^{\rm BSM}_{\rm eff} \sim \( \frac{\Lambda_\chi}{\Lambda_{\rm BSM}} \)^2 \sim 10^{-10} 
\( \frac{100 \ \text{TeV}}{\Lambda_{\rm BSM}} \)^2 ~ , 
\eeq
where in the second step we have normalized $\theta^{\rm BSM}_{\rm eff}$ to the bound set by the nEDM. 
This expression suggests that the QCD axion could act as a low-energy probe  
of high-energy sources of CP-violation, by testing generic CP-violating new physics up to the scale $\Lambda_{\rm BSM} \lesssim 100$ TeV. 

An axion VEV could also be generated from an ultraviolet (UV) 
source of PQ breaking that is not aligned to the QCD anomaly.  This is motivated by the fact that the U(1)$_{\rm PQ}$,  
being a global symmetry, does not need to be exact and it is expected to be broken by quantum gravity effects 
(see e.g.~\cite{Kallosh:1995hi}). 
Hence, the axion VEV could be displaced from zero, leading to the so-called PQ 
quality problem \cite{Barr:1992qq,Holman:1992us,Kamionkowski:1992mf}, that is the need to explain why UV physics does not generically spoil the QCD axion solution 
to the strong CP problem. There actually exist UV constructions, based e.g.~on extra gauge symmetries 
(see e.g.~\cite{Randall:1992ut,Dobrescu:1996jp,Redi:2016esr,Fukuda:2017ylt,DiLuzio:2017tjx,Bonnefoy:2018ibr,Gavela:2018paw,Ardu:2020qmo,DiLuzio:2020qio,Vecchi:2021shj,Contino:2021ayn}), 
which ensure a controlled solution to the PQ quality problem. In the following, we will assume as a 
canonical example a PQ-breaking 
effective
operator of the type 
\beq
\label{eq:PQbreak}
V_{\text{PQ-break}}= e^{i \delta_\theta} \frac{\Phi^n}{\Lambda_{\rm UV}^{n-4}}+\text{h.c.} ~ , 
\eeq
with $n>4$. Here, $\delta_\theta$ denotes a generic phase,   
$\Lambda_{\rm UV}$ is some UV scale like the Planck mass 
and $\Phi=\frac{f_a}{\sqrt{2}} e^{i\theta} $ is a complex scalar,  
with the radial mode integrated out and 
the angular mode $\theta = a /f_a $ given by the axion. 
In terms of $\theta$, the PQ-breaking potential reads
\begin{gather}
V_{\text{PQ-break}}= \frac{1}{2^{n/2-1}} \frac{f_a^n}{\Lambda_{\rm UV}^{n-4}}\cos(n\theta+\delta_{\theta}) ~ .\label{eq:cosine from higher dim}
\end{gather}
In general, the phase of this potential, parameterized by $\delta_{\theta}$, needs not be aligned with the CP-conserving minimum of the QCD potential at $\theta=0$. Therefore, we 
generically expect $\delta_{\theta}$ to be $\mathcal{O}(1)$.
The balance between $ V_{\rm PQ-break} $ and the standard QCD axion potential, $V_{\rm QCD} = \frac{1}{2} f_a^2 m_a^2 \theta^2 + \mathcal{O}(\theta^4)$, 
yields the axion VEV (approximated for $\theta\ll 1$)  
\begin{gather}
	\theta_{\rm eff} 
	\simeq - \frac{n \Lambda_{\rm UV}^{4-n} f_a^n \sin \delta_\theta}{2^{\frac{n}{2}-1} f_a^2 m_a^2} ~ , 
\end{gather}
which effectively also breaks CP. 
Taking for instance $\sin\delta_\theta = 1$, 
$\Lambda_{\rm UV} = M_{\rm Pl}$ and 
$f_a = 10^{8}$ 
GeV, 
the nEDM bound $|\theta_{\rm eff}| \lesssim 10^{-10} $ 
translates into $n\geq 9$. 

While the 
nEDM directly probes 
$\theta_{\rm eff}$, 
the presence of a propagating light axion field 
could offer additional ways to test the axion ground state, 
as discussed in the following.  

\subsection{CP-violating axion-nucleon couplings}
\label{sec:masterformula}

A remarkable consequence of $\theta_{\rm eff} \neq 0$ is the 
generation of a \emph{scalar} axion coupling to nucleons, 
$g^S_{aN} a \bar N N$ (with $N= p, n$), which can be searched 
for in axion-mediated force experiments as suggested by 
Moody and Wilczek \cite{Moody:1984ba}.  
The phenomenological relevance of these observables
will be discussed in \sect{sec:axionforces}. 
To understand the origin of $g^S_{aN}$, let us define the canonical axion field as the 
excitation over its VEV, 
$a \to \theta_{\rm eff} f_a + a$, and consider the two-flavour QCD axion Lagrangian 
\beq 
\label{eq:LaQCD}
\mathcal{L}_a = \frac{\alpha_s}{8\pi} 
\( \theta_{\rm eff} + \frac{a}{f_a} \) G \tilde G - \bar q_L M_q q_R + \text{h.c.} ~ , 
\eeq
where $M_q = \diag (m_u, m_d)$. 
Upon the quark field redefinition 
\beq 
\label{eq:qaxial}
q \to e^{\frac{i}{2}\gamma_5 \( \theta_{\rm eff} + \frac{a}{f_a} \) Q_a} q ~ ,
\eeq
with $Q_a = M_q^{-1} / \Tr M_q^{-1} = \diag (\frac{m_d}{m_u+m_d}, \frac{m_u}{m_u+m_d}) $, 
the $G\tilde G$ term in \eq{eq:LaQCD} gets rotated away, while the 
quark mass matrix becomes 
\beq 
\label{eq:Ma}
M_q \to M_a = e^{i\gamma_5 \( \theta_{\rm eff} + \frac{a}{f_a} \) Q_a} M_q ~ , 
\eeq
where we used $[M_q, Q_a] = 0$. Hence, in the new basis we have 
\begin{align}
\label{eq:LaQCD2}
\mathcal{L}_a &= - \bar q \[ \frac{M_a + M^\dag_a}{2} 
+ \gamma_5 \frac{M_a - M^\dag_a}{2} \] q \nonumber \\
&= - \bar q  \cos \( \( \theta_{\rm eff} + \frac{a}{f_a} \) Q_a \) M_q q
- \bar q  i \gamma_5 \sin \( \( \theta_{\rm eff} + \frac{a}{f_a} \) Q_a \) M_q q
~ . 
\end{align}
Focusing on the scalar component (i.e.~the term without $\gamma_5$) 
\begin{align}
\mathcal{L}_a &\supset 
- \bar q \cos \( \theta_{\rm eff} Q_a \) \cos \( \frac{a}{f_a} Q_a \)  M_q q
+ \bar q \sin \( \theta_{\rm eff} Q_a \) \sin \( \frac{a}{f_a} Q_a \) M_q q \nonumber \\ 
&\simeq \theta_{\rm eff}  \frac{a}{f_a} \bar q ~ Q_a^2  M_q q = 
\frac{\theta_{\rm eff}}{f_a} \frac{m_u m_d}{m_u + m_d} a 
\( \frac{m_d}{m_u + m_d} \bar u u + \frac{m_u}{m_u + m_d} \bar d d \) 
~ ,  
\end{align}
where in the second step we kept only a term linear in $a/f_a$ and expanded for small $\theta_{\rm eff}$, 
while in the last step we employed the explicit representation of $Q_a$ given below \eq{eq:qaxial}. 
Hence, the scalar axion coupling to nucleons is 
\begin{align} 
g^S_{aN} &= \frac{\theta_{\rm eff}}{f_a} \frac{m_u m_d}{m_u + m_d} 
\( \frac{m_d}{m_u + m_d} \langle N | \bar u u | N \rangle + \frac{m_u}{m_u + m_d} \langle N | \bar d d | N \rangle \)  \\
&= \frac{\theta_{\rm eff}}{f_a} \frac{m_u m_d}{m_u + m_d}
\[ 
\frac{\langle N | \bar u u + \bar d d | N \rangle}{2}
+ \( \frac{m_d - m_u}{m_u + m_d} \) \frac{\langle N | \bar u u - \bar d d | N \rangle}{2} \] \nonumber ~ ,  
\end{align}
where in the last step we isolated the iso-spin singlet component, 
$\bar u u + \bar d d$.   
Focusing on the leading iso-spin singlet term (as in \cite{Moody:1984ba}), one has  
\beq 
\label{eq:fromthetaefftogaN}
g^S_{aN} \simeq \frac{\theta_{\rm eff}}{f_a} \frac{m_u m_d}{m_u + m_d} 
\frac{\langle N | \bar u u + \bar d d | N \rangle}{2} 
\simeq 1.3 \cdot 10^{-12} ~ \theta_{\rm eff} \( \frac{10^{10} \ \text{GeV}}{f_a} \) 
~ . 
\eeq 
For the numerical evaluation, we employed the 
value 
of the 
pion-nucleon sigma term 
$\sigma_{\pi N} = \langle N | \bar u u + \bar d d | N \rangle (m_u + m_d) / 2 = 59.1 \pm 3.5$ MeV \cite{Hoferichter:2015dsa}. 
As pointed out in Ref.~\cite{Bertolini:2020hjc},  
a factor $1/2$ was missed 
in the original derivation of Moody and Wilczek \cite{Moody:1984ba}. 

Note that \eq{eq:fromthetaefftogaN} is formally correct 
only in the case in which the origin of the axion VEV 
is due to a UV source of PQ breaking, as e.g.~in \eq{eq:PQbreak}. 
On the other hand, in the more general case of 
CP-violating sources related to QCD, i.e.~a short distance 
operator made of quarks and gluons,  
\eq{eq:fromthetaefftogaN} turns out to be conceptually 
unsatisfactory when we focus on the relevant question: \emph{how to properly 
correlate the scalar axion-nucleon coupling with the nEDM bound?} 
The reason being the \eq{eq:fromthetaefftogaN} misses extra contributions due to meson tadpoles ($\pi^0$, $\eta$, $\eta'$), 
which are 
generated by the same UV source of CP violation responsible for $\theta_{\rm eff} \neq 0$
and  
are of the same size of the term proportional to $\theta_{\rm eff}$. 
This improvement was recently taken into account in Ref.~\cite{Bertolini:2020hjc} which,  
including as well iso-spin breaking effects from the leading order axion-baryon-meson chiral Lagrangian, 
found the master formula 
\begin{align}
\label{gan-pVEV}
{g}^S_{an,\,p} &\simeq 
\frac{4B_0\, m_u m_d}{f_a (m_u+m_d)}  \bigg[\pm (b_D+b_F)\frac{\vev{\pi^0}}{F_\pi}
+ \frac{b_D-3b_F}{\sqrt{3}}\frac{\vev{\eta_8}}{F_\pi}  
 -\sqrt{\frac{2}{3}}(3b_0+2b_D) \frac{\vev{\eta_0}}{F_\pi} \nonumber \\
&-   \left(b_0 + (b_D+b_F)\frac{m_{u,d}}{m_d+m_u}\right)\theta_{\rm eff}\bigg] ~ , 
\end{align}
where for clarity we neglected $m_{u,d}/m_s$ terms. Here, $B_0=m_\pi^2/(m_d+m_u)$ while the hadronic
Lagrangian parameters $b_{D,F}$ are determined from the baryon octet mass splittings,
$b_D\simeq 0.07\,\rm GeV^{-1}$, $b_F\simeq -0.21\,\rm GeV^{-1}$ at LO~\cite{Pich:1991fq}.  The value
of $b_0$ is determined from the pion-nucleon sigma term as $b_0\simeq -\sigma_{\pi N}/4m_\pi^2$.
From the determination in~\cite{Hoferichter:2015dsa} one obtains
$b_0\simeq -0.76\pm 0.04\, \rm GeV^{-1}$. 
Given $\sigma_{\pi N}\equiv \langle N | \bar u u + \bar d d | N \rangle\, (m_u+m_d)/2$, 
the isospin symmetric  $b_0\theta_{\rm eff}$ term reproduces exactly \eq{eq:fromthetaefftogaN}.

In general, $g^S_{aN}$ and $d_n$ are \emph{not} proportional, as it would follow instead 
 from~\eq{eq:fromthetaefftogaN}. For instance, exact cancellations among the VEVs can happen
for $d_n$~\cite{Cirigliano:2016yhc,Bertolini:2019out} which have no counterpart in $g^S_{aN}$.
Note that $\theta_{\rm eff}$ and the meson VEVs 
in \eq{gan-pVEV}  
are meant to be computed from high-energy sources of
CP violation, represented by an effective operator $\mathcal{O}_{\rm CPV}$. 
In Ref.~\cite{Bertolini:2020hjc} an explicit example 
was worked out in the context of 4-quark operators of the type 
$\mathcal{O}_{\rm CPV} = (\bar q q) (\bar q' i \gamma_5 q')$ 
with $q,q'= u,d,s$, 
arising e.g.~by integrating out the heavy $W_R$ boson in 
left-right symmetric models. Building 
on the detailed analysis of Ref.~\cite{Bertolini:2019out},  
both $g^S_{aN}$ and $d_n$ were computed in the minimal left-right symmetric model 
with $\mathcal{P}$-parity \cite{Senjanovic:1978ev,Mohapatra:1979ia}, 
showing a non-trivial interplay which deviates sizeably from the naive scaling implied by 
\eq{eq:fromthetaefftogaN}, that is $g^S_{aN} \propto d_n \propto \theta_{\rm eff}$. 

More recent works have readdressed the calculation of $g^S_{aN}$ in terms of short-distance 
sources of CP violation. Ref.~\cite{Okawa:2021fto} considered the contribution of 
EDMs and color EDMs of quarks, and revisited as well the SM contribution arising from the CKM phase. 
Ref.~\cite{Dekens:2022gha} provided instead a systematic calculation of CP-violating axion couplings, 
starting from the CP-violating basis of low-energy EFT (LEFT) 
operators of quarks and gluons, and employing at low-energy 
a chiral Lagrangian approach, similar to the one of \eq{gan-pVEV}. 

\subsection{Axion-mediated forces vs nEDM}
\label{sec:axionforces}

Including both scalar and pseudo-scalar couplings to matter fields,  
the axion interaction Lagrangian  can be written as
\beq 
\label{eq:Laint}
\mathcal{L}^{\rm int}_a \supset 
g^S_{aN} a \bar N N + 
g^P_{af} a \bar f i \gamma_5 f ~ ,  
\eeq
where $f = N, e$, and $g^P_{af} = C_f m_f / f_a$ denoting the usual pseudo-scalar axion coupling, 
with $C_f  \sim \mathcal{O}(1)$ in benchmark axion models \cite{DiLuzio:2020wdo}. 

By taking the non-relativistic limit of $\mathcal{L}^{\rm int}_a$ one obtains different kinds of static potentials,  
which can manifest themselves as new axion-mediated macroscopic forces \cite{Moody:1984ba}. 
The latter can be tested in laboratory experiments, and hence do not rely on model-dependent axion production mechanisms, 
as in the case of dark matter axions (haloscopes) or to a lesser extent solar axions (helioscopes). 
An updated review of axion-mediated force experiments and relevant limits can be found in 
Ref.~\cite{OHare:2020wah} (see also \cite{Raffelt:2012sp,Irastorza:2018dyq,Sikivie:2020zpn}). 

Axion-induced potentials can be of three types,  
depending on the combination of couplings involved, 
i.e.~$g^S_{aN} g^S_{af}$ (monopole-monopole),
$g^S_{aN} g^P_{af}$ (monopole-dipole)
and $g^P_{af} g^P_{af}$ (dipole-dipole). 
The idea of searching for dipole-dipole axion interactions   
in atomic physics 
is as old as the axion itself \cite{Weinberg:1977ma}. 
However, dipole-dipole forces turn out to be 
spin suppressed in the non-relativistic limit 
and suffer from large backgrounds from ordinary magnetic forces. 
Searches based on monopole-monopole interactions, 
like tests of gravity on macroscopic scales, are in principle much more powerful.  
However, under the theoretical prejudice that we are after the QCD axion, 
the $\theta_{\rm eff}^2$ suppression (given the nEDM bound) is such that 
current experiments are still some orders of magnitude away from testing the QCD axion.  

\begin{figure}[t]
\centering
\includegraphics[width=15cm]{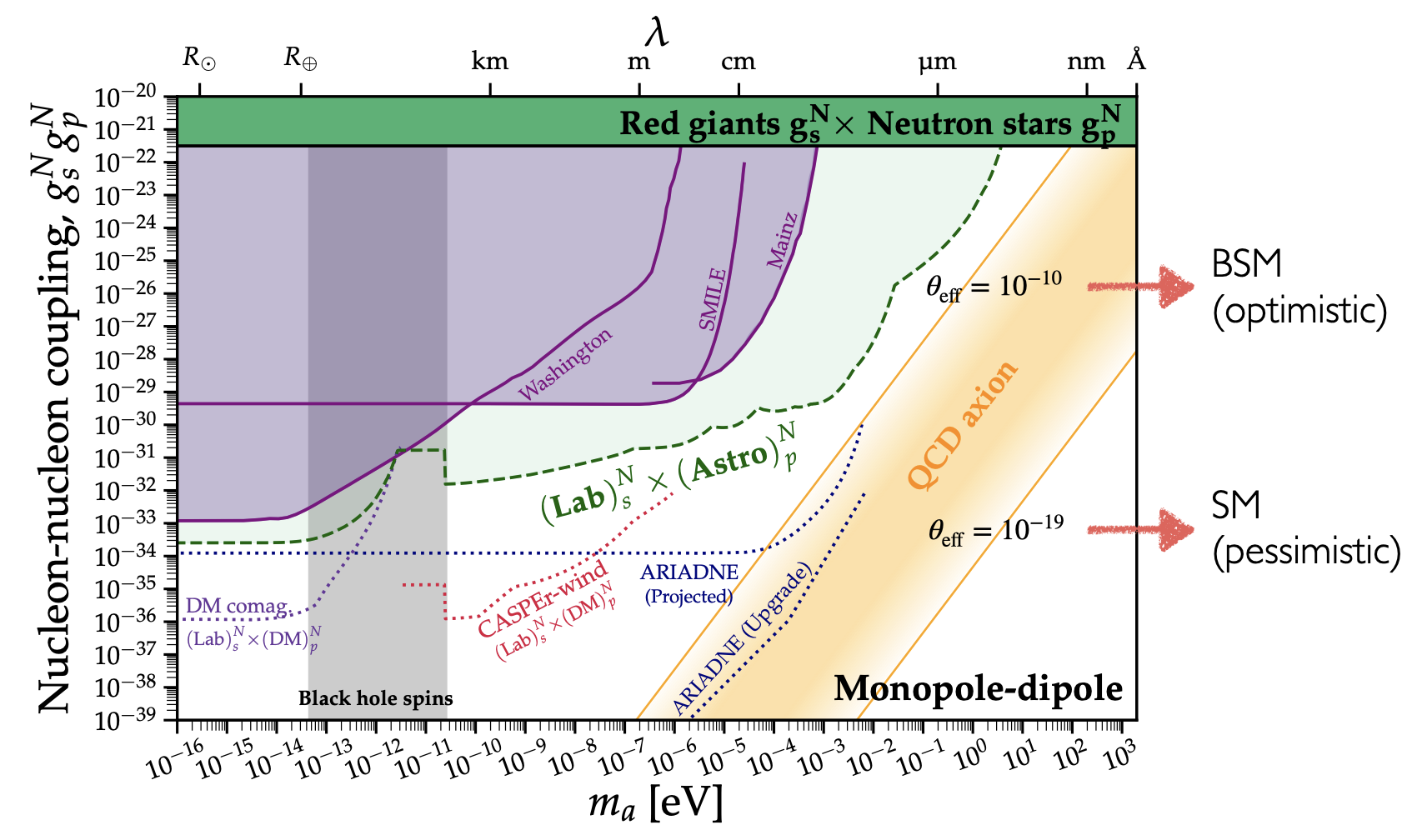}
\caption{Parameter space of axion-mediated monopole-dipole forces. Figure adapted from Ref.~\cite{OHare:2020wah}.}
\label{fig:mondip}       
\end{figure}
The sweet spot is given by monopole-dipole searches which, as shown in \fig{fig:mondip}, 
will enter into the QCD axion region. 
In this case, the axion-mediated potential between nucleons and a spin sample of fermions of type $f$ 
is given by \cite{Moody:1984ba}
\beq 
\label{eq:mondip}
V(r) = g^S_{aN} g^P_{af} \frac{\hat{\sigma} \cdot \hat{r}}{8\pi M_f} \( \frac{m_a}{r} + \frac{1}{r^2} \) e^{-m_a r} ~ ,
\eeq 
which implies that for $m_a r \lesssim 1$, so that the potential is not exponentially suppressed, 
$V(r)$ falls like $1/r^2$. 
A new detection concept by the ARIADNE collaboration \cite{Arvanitaki:2014dfa,Geraci:2017bmq}
plans to use nuclear magnetic resonance techniques to probe the axion field sourced by an unpolarized material 
via a sample of nucleon spins.\footnote{A similar approach is pursued by the QUAX-$g_pg_s$ 
collaboration \cite{Crescini:2016lwj,Crescini:2017uxs} using instead electron spins.} 
Note that the yellow QCD axion band in \fig{fig:mondip} is obtained by 
employing \eq{eq:fromthetaefftogaN} for the scalar axion-nucleon coupling, 
with the value of $\theta_{\rm eff}$ spanning from the SM estimate 
$\theta_{\rm eff} \sim 10^{-19}$ (lower side)
to the limit imposed by the nEDM  
$\theta_{\rm eff} \simeq 10^{-10}$ (upper side). 
However, as argued in \sect{sec:masterformula}, the relation between $g^S_{aN}$ and $d_n$ 
is model dependent and hence, given a specific source of CP-violation,  
it should be assessed case by case in order to properly determine the parameter space region 
that is allowed by the nEDM bound (for a comprehensive analysis, see Ref.~\cite{Dekens:2022gha}). 

In order to have a testable signal in monopole-dipole axion searches 
a sizeable source of CP violation beyond the SM is hence required. 
As a reference value in terms of $\theta_{\rm eff}$ 
(using \eq{eq:fromthetaefftogaN} for $g^S_{aN}$) 
one would need at least 
$\theta_{\rm eff} \gtrsim 10^{-13}$, 
in order for this effect to be testable at the ultimate phase of 
ARIADNE (cf.~\fig{fig:mondip}). 
While the SM prediction, $\theta_{\rm eff} \sim 10^{-19}$, 
is 
far from being testable,  
new sources of CP violation 
beyond the CKM phase
are needed to explain the 
matter-antimatter asymmetry of the universe. 
Even if they are
decoupled 
at very high-energy scales,  
these CP-violating sources might contribute sizeably to $\theta_{\rm eff}$, 
as shown by the estimate in \eq{eq:thetaeffmin3}, 
and could become accessible in axion-mediated force experiments. 

Future improvements 
of EDM limits 
will also play a crucial role in disentangling 
different sources of CP violation coupled to QCD, 
also in the presence of a light axion field (see e.g.~\cite{deVries:2021sxz,Choi:2023bou}). 
From this perspective, the interplay of low-energy CP-violating phenomena involving the axion field, 
such as axion-mediated forces and EDMs, 
offers a new window for testing high-energy sources of CP-violation.

\subsection{Cosmological signatures of CP-violating axions}
\label{sec:CPVaxioncosmo}

\newcommand{\Tosc}{T_{\rm osc}}

Today, CP-violating axion potentials are significantly limited
by the nEDM constraint, making them strongly subdominant compared to the axion potential from QCD. However, the potential generated by QCD in the form of the topological susceptibility 
$K(T)$, defined in Eq.~\eqref{eq:defK}, 
becomes strongly suppressed at temperatures above the QCD crossover. Lattice QCD determinations of $K(T)$,
such as \cite{Borsanyi:2016ksw}, can be approximately 
fitted with the analytical expression 
\begin{gather}
    \frac{K(T)}{f_a^2} = m_a^2(T) = m_a^2 \, \max\left[ \left(\frac{T}{T_c}\right)^{-2b} ,\ 1\right]\label{eq:ma(T) lattice fit} ~ , 
\end{gather}
where $T_c = 150 $ MeV and $b=3.92$. At early times, the QCD axion potential is subdominant to Hubble friction and the field remains frozen at a constant value. Later, once the temperature $\Tosc$ is reached where the condition $m_a(\Tosc)\approx 3 H(\Tosc)$ is satisfied, the field starts oscillating and the evolution can be described in the WKB approximation as $a \approx m_a^{-1/2}(T)\ R^{-3/2}$, 
where $R$ denotes the cosmological scale factor. For an axion with no other significant potential, this leads to a dark matter relic density of the form
\begin{gather}
	\rho_{a,\,\rm today} \simeq \frac{1}{2}m_a^2 f_a^2\delta^2 \frac{m_a(\Tosc)}{m_a}\frac{g_{*s}(T_{\rm today})}{g_{*s}(\Tosc)}\left(\frac{T_{\rm today}}{\Tosc}\right)^3~,
\end{gather}
where we neglected any anharmonic effects, which contribute an $\mathcal{O}(1)$ factor unless $\delta$ is tuned to $\pi$, and made the simplifying assumption that the field can be immediately described by the WKB approximation once $m_a(T)\approx 3 H(T)$ is satisfied. Here, $\delta \in [-\pi,\pi]$ is the initial misalignment angle, which we generically expect to be $\mathcal{O}(1).$ To reproduce a fraction $ r\equiv \rho_{a,\,\rm today} / \rho_{\rm DM,\, today}$, this corresponds to an oscillation temperature of
\begin{gather}
    \Tosc \simeq 900 \text{ MeV } \times  r^{-1/7} \delta^{2/7} ~,
\end{gather}
where we approximated $b=4$ to simplify the functional form. This implies that the CP-preserving potential is suppressed relative to the present-day value by  
\begin{gather}
    \frac{K(\Tosc)}{K} \simeq 5\times 10^{-7}\times  r^{8/7}\delta^{-16/7} \label{eq:K suppression}~.
\end{gather}
Assuming an $\mathcal{O}(1)$ misalignment angle and that the axion makes up the whole dark matter density thus implies that the CP-preserving potential $\propto K(T) $ is suppressed by a factor of more than $10^6$ relative to the present day value. Therefore, although any CP-violating potential must be subdominant today, it is conceivable that it played a relevant role at temperatures at or above the QCD crossover. 

The possibility for a CP-violating potential to affect the 
standard misalignment mechanism crucially depends on 
the potential's variation with temperature. If the CP-violating effect is sourced 
by an effective operator, $\mathcal{O}_{\rm CPV}$, 
coupled to QCD, as in \eq{eq:defKp}, then we expect 
$K'$ to be power-suppressed at $T \gtrsim T_c$ 
in a similar way as for the topological susceptibility.
However, if 
the CP-violating potential does not depend on the temperature, as in the case of a UV source of PQ-breaking exemplified by Eq.~\eqref{eq:PQbreak}, then it may become competitive with the topological susceptibility at temperatures relevant for misalignment. 

In general, 
a CP-violating potential, parameterized by $K'$ (see \eq{eq:Vthetaeff}), dominates the misalignment dynamics if $K'(T)  \gtrsim K(T)$. Today, this hierarchy is limited by 
the nEDM to be  $K'/K \lesssim 10^{-10}$. If $K'$ is constant with temperature, then $K' \gtrsim K(\Tosc)$ requires 
\begin{gather}
\frac{K(\Tosc)}{K} \lesssim 10^{-10}~.
\end{gather}
From Eq.~\eqref{eq:K suppression} we observe that the assumptions of an $\mathcal{O(1)}$ misalignment angle and $r=1$, i.e.~that axions make up all of the dark matter, implies that the CP-violating potential remains subdominant at $\Tosc$. However, in the regime where the axion makes up only a sub-fraction of dark matter, $K' \gtrsim K(\Tosc) $ can be achieved for $r\lesssim 5 \times 10^{-4}$. This implies that 
CP-violating potentials significantly impact misalignment for axions with
\begin{gather}
\label{eq:maa}
m_a \gtrsim 5 \times 10^{-3} \text{ eV} ~,
\end{gather}
where we still assumed $\delta_\theta = \mathcal{O}(1)$. 
We arrived at this conclusion by starting with the assumption that the CP-preserving potential dominates misalignment, so a more careful analysis is required in the regime where the CP-violating potential is dominant.

The regime in which the CP-violating potential significantly modifies the relic abundance was studied in Ref.~\cite{Jeong:2022kdr}. In their work, the authors considered how the evolution generated by the temperature-dependent axion mass, given by the fit in \eq{eq:ma(T) lattice fit}, is modified by the presence of a constant-temperature CP-violating potential similar to \eq{eq:cosine from higher dim}. 
For such a potential, there are two regimes with distinct phenomenology, which depend on the phase of the CP-violating potential $\delta_\theta$ and the misalignment angle of the axion $\delta$:
\begin{align*}
    \text{Smooth shift regime:} \qquad \abs{\delta - \delta_\theta } &< \pi / n ~, \\
  \text{Trapping regime:} \qquad \abs{\delta - \delta_\theta } &> \pi / n ~, 
\end{align*}
where $n$ denotes the number of minima of the axion potential, 
see \eq{eq:cosine from higher dim}. 
In either case, the axion initially rolls towards the nearest minimum of the CP-violating potential before the time when it would have started oscillations around the CP-preserving minimum. Eventually, the axion will transition to the 
CP-preserving minimum as this becomes dominant. The crucial distinction is whether the transition between these minima is adiabatic (smooth shift) or if it takes place abruptly (trapping). The smooth shift will take place if the axion initially oscillates in a CP-violating minimum 
neighbouring the CP-preserving one. In this case, oscillations start in the CP-violating minimum and continue to decay according to the conventional WKB evolution throughout the transition.
The effect is then to decrease the relic abundance by triggering earlier oscillations and thus increasing the redshift. 

Alternatively, if the axion initially starts oscillating in a CP-violating minima that does not neighbour the CP-preserving one, then there will be a potential barrier preventing the transition from taking place in an adiabatic way. Instead, at the transition the axion will be suddenly released from the higher CP-violating minimum and new oscillations will be triggered with an amplitude set by the misalignment between the false and true minima. This mechanism, known 
in the literature as 
trapped misalignment, has been discussed in different 
contexts, see e.g.~\cite{Higaki:2016yqk,Kawasaki:2017xwt,Nakagawa:2020zjr,DiLuzio:2021gos}.  
Interestingly, in this limit, the axion 
abundance becomes approximately independent of $m_a$ 
(up to changes in $g_{*s}$), 
and the present-day relic is controlled entirely by the position of the trapped minimum and the amplitude of the CP-violating potential. 
Thus, the dark matter abundance can either be increased or decreased by the CP-violating potential. However, as argued above, if there is no tuning then the CP-violating potential can only become significant in the regime in which misalignment  under-produces axion dark matter, cf.~\eq{eq:maa}. 
The most optimistic scenario is thus that the CP-violating potential could enhance the fraction of dark matter that the axion can make up in the $m_a \gtrsim 10^{-3}$ eV regime. 
Since 
the axion abundance is approximately 
independent of $m_a$, the net effect is that the misalignment relic produced at larger $m_a$ would be similar to the one at the transition point.

The reason why the CP-violating potential cannot generically modify the axion dark matter relic density is because of the nEDM constraint. However, if the phase of the CP-violating potential $\delta_\theta$ is tuned towards zero, then the nEDM constraint can be relaxed circumventing the limitations above. Ref.~\cite{Jeong:2022kdr} finds that with sufficient tuning,
the CP-violating potential in \eq{eq:cosine from higher dim} can reproduce the observed dark matter from misalignment for any $m_a$ compatible with observational constraints. For $m_a \gg 10^{-5}$ eV, which conventionally under-produces axion dark matter, the axion abundance can be enhanced through trapping. For $m_a \ll 10^{-5}$ eV, which conventionally over-produces axion dark matter for $\delta\sim \mathcal{O}(1)$ misalignment angles, the axion relic can be diluted if the axion starts in the smooth shift regime. In the latter scenario, tuning in $\delta_\theta$ replaces tuning in $\delta$.

\section{CP-violating ALPs}
\label{sec:ALPs}

The ALP, hereafter denoted by $\phi$, is a popular generalization of the QCD axion discussed in \sect{sec:QCDaxions}, with the most prominent difference being that its mass ($m_\phi$) and couplings to SM particles are arbitrary parameters to be determined or bounded by experiments.
Here, we only require that $m_\phi \ll \Lambda$, where $\Lambda$ is a UV scale related to the origin of the ALP field. 
%
In the following, we report on some recent results from Refs.~\cite{DiLuzio:2020oah,DiLuzio:2023edk} regarding the 
construction of the most general CP-violating ALP interactions 
and related contributions to different types of EDMs. 
The treatment differs depending on whether the ALP mass is 
above the GeV scale \cite{DiLuzio:2020oah}, as detailed in \sect{sec:ALPabove1GeV}, or below it \cite{DiLuzio:2023edk}, 
as further discussed in \sect{sec:ALPbelow1GeV}. 

\subsection{CP-violating ALP Lagrangian: $m_\phi \gtrsim 1$ GeV} 
\label{sec:ALPabove1GeV}
Let us start by considering the case $m_\phi \gtrsim \rm few~$GeV, so that QCD can be treated perturbatively. Moreover, although we take $\Lambda\gtrsim 1$ TeV, we focus only on electromagnetic and strong interactions, since weak interactions play a subleading role in our analysis. 

The standard ALP effective Lagrangian \cite{Georgi:1986df} 
can be generalised to include also CP-violating interactions.
The most general $SU(3)_c \times U(1)_{\rm em}$ invariant $d=5$ effective Lagrangian below the electroweak scale, 
accounting for CP-violating ALP interactions 
with photons, gluons and SM fermions, reads \cite{DiLuzio:2020oah}
\begin{align} 
\label{eq:Lphi}
\mathcal{L}_{\phi} & \supset  
e^2 \frac{\tilde C_{\gamma}}{\Lambda} \phi F \tilde F + g^2_s \frac{\tilde C_{g}}{\Lambda} \phi G \tilde G  + i \frac{v}{\Lambda} y^{ij}_P \phi \bar f_i \gamma_5 f_j 
\nonumber\\
& + e^2 \frac{C_{\gamma}}{\Lambda} \phi F F + g^2_s \frac{C_{g}}{\Lambda}  \phi G G  
+ \frac{v}{\Lambda} y^{ij}_S \phi \bar f_i f_j ~ , 
\end{align}
where $f \in (e,u,d)$ denotes SM fermions in the mass basis, $i,j$ are flavor indices, and $y_{S}$ and $y_{P}$ are hermitian matrices.
$F$ and $G$ stand for the QED and QCD field strength tensors, respectively, while $\tilde F$ and $\tilde G$ are their duals. 
Contractions in Lorentz space are implicitly assumed. 
Note that the interactions in the first line of Eq.~(\ref{eq:Lphi}) can be understood to be invariant under the $\phi$ shift symmetry, as long as 
non-perturbative effects related to boundary terms can be neglected in the case of QCD. 
In fact, both $F \tilde F$ and $G \tilde G$ are total derivatives, while pseudo-scalar interactions 
can be understood as the linear expansion of an exponential parametrization of the ALP field.  
The latter interaction can be rewritten in a shift-symmetric way through the $d=5$ operator $\frac{\partial_\mu \phi}{\Lambda} \bar f_i \gamma^\mu \gamma_5 f_j$, 
upon an anomalous ALP-dependent field redefinition on the fermion fields, which  
also redefines the couplings $\tilde C_{\gamma}$ and $\tilde C_{g}$.\footnote{Note that the identification between the 
pseudo-scalar operator in \eq{eq:Lphi} and the derivative ALP interaction with the axial current breaks down for processes involving the exchange of more than one ALP field. 
In the following, we will focus exclusively on processes that entail the exchange of a single ALP field. Therefore, the parameterization provided in \eq{eq:Lphi} suffices for our purposes.}  
Instead, the interactions in the second line of Eq.~(\ref{eq:Lphi}) break explicitly the ALP shift symmetry. In particular, scalar interactions can be written, in the unbroken SM phase, in terms of the $d=5$ operator $\phi H\bar f_{L(R)} f_{R(L)}$ thus justifying the normalization factor $v/\Lambda$, with $v=246$ GeV. 
Notice that $(\tilde{C}_\gamma, \tilde{C}_g, y_P)$ and $(C_\gamma, C_g, y_S)$ could have very different sizes as they stem from the shift-symmetry invariant and breaking sectors, respectively.
For a detailed account of the shift-breaking orders in the ALP effective Lagrangian, see Ref.~\cite{Bonnefoy:2022rik}. 
Possible UV completion of the CP-violating ALP effective Lagrangian in \eq{eq:Lphi} will be discussed in 
\sect{sec:UVALP}. 

Since both the operators $XX$ and $X \tilde X$ ($X= F,G$), and scalar and pseudo-scalar operators have opposite CP transformation properties, $\mathcal{L}_{\phi}$ necessarily violates CP irrespectively of the CP-even or CP-odd nature of $\phi$.
CP-violating phenomena are commonly described by means 
of Jarlskog invariants,\footnote{See 
Ref.~\cite{Bonnefoy:2022rik} for 
a classification of the Jarlskog invariants at the SM EFT level.} 
i.e.~rephasing-invariant parameters which provide a measure of CP violation~\cite{Jarlskog:1985ht}. 
In our ALP scenario, the full set of Jarlskog invariants is given by
\begin{align} 
\label{eq:Jarlskog}
C_{a}\tilde C_{b} ~, \quad y^{ii}_S \, {\tilde C_{a}} ~, \quad 
y^{ii}_P \, C_{a} \,, \quad y^{ii}_S \, y^{jj}_P \,, \quad y^{ik}_S \, y^{kk}_{\rm SM} \, y^{ki}_P\,,
\end{align}
where 
$a,b = \gamma,g$ 
and $y^{kk}_{\rm SM}$ denotes a SM Yukawa coupling in the diagonal basis. Notice that only the last invariant of Eq.~(\ref{eq:Jarlskog}) is sensitive to flavor-violating effects.  

\subsubsection{ALP contribution to EDMs} 

\begin{figure}
\centering
\includegraphics[width=1.0\linewidth]{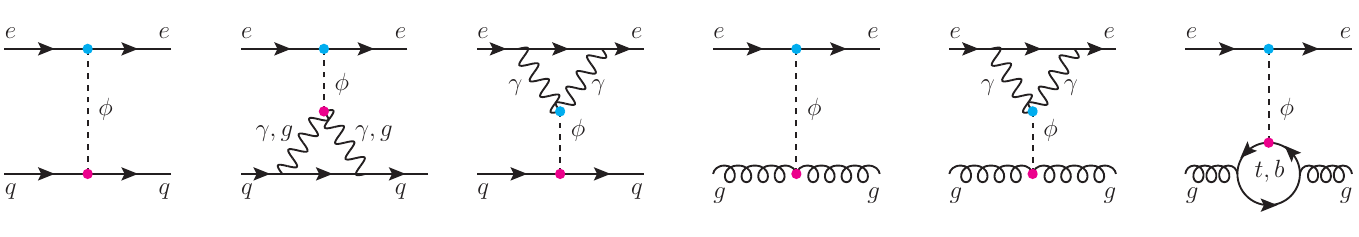}
\caption{Leading contributions to the semi-leptonic nucleon-electron operators.
The combination of light-blue and purple blobs refer to CP-violating effective interaction vertices. 
Figure from Ref.~\cite{DiLuzio:2020oah}.}
\label{fig:4f}
\end{figure}
%
The effective Lagrangian $\mathcal{L}_{\phi}$ 
defined in \eq{eq:Lphi} at the scale $\Lambda$ is renormalized 
at lower energies by QED and QCD interactions.\footnote{The complete one-loop anomalous dimension matrix assuming the full SM gauge group has been 
recently presented in Ref.~\cite{DasBakshi:2023lca}.}
In the leading logarithmic approximation, 
the solution of the renormalization-group equations for the leptonic (pseudo)scalar couplings $y^{\ell\ell}_{S,P}$ 
gives~\cite{DiLuzio:2020oah}
\begin{align}
\label{eq:y_l_RGE}
y^{\ell\ell}_S & \simeq y^{\ell\ell}_S(\Lambda) + \frac{6\alpha}{\pi} \frac{m_\ell}{v} e^2 C_\gamma \log\frac{\Lambda}{\mu}~,
\\
y^{\ell\ell}_P & \simeq y^{\ell\ell}_P(\Lambda) - \frac{6\alpha}{\pi} \frac{m_\ell}{v} e^2 {\tilde C}_\gamma \log\frac{\Lambda}{\mu}~,
\end{align}
where $\mu$ is the renormalization scale. Instead, in the quark sector, we obtain~\cite{DiLuzio:2020oah}
\begin{align}
\!\! y^{qq}_S \simeq & ~~ y^{qq}_S(\Lambda) \!+\! 
\frac{m_q}{v} \! \left( \frac{6\alpha}{\pi} Q^2_q e^2 C_\gamma \!+\! \frac{8\alpha_s}{\pi} g_s^2 C_g \right) \!
\log\frac{\Lambda}{\mu}~,
\\
\! y^{qq}_P \simeq & ~~ y^{qq}_P(\Lambda) \!-\! 
\frac{m_q}{v} \! \left( \frac{6\alpha}{\pi} Q^2_q e^2 {\tilde C}_\gamma \!+\! \frac{8\alpha_s}{\pi} g_s^2 {\tilde C}_g \right)\!
\log\frac{\Lambda}{\mu}~.
\label{eq:y_q_RGE}
\end{align}
Since in Eq.~(\ref{eq:Lphi}) we factored out the gauge couplings $e^2$ and $g^2_s$, the coefficients $C_{\gamma,g}$ and $\tilde C_{\gamma,g}$ are
scale-invariant at one-loop order.
Top and bottom contributions are taken into account by the QCD trace anomaly in the gluon-gluon-ALP vertex after they 
have been integrated out (see Fig.~\ref{fig:4f}), yielding 
\begin{align}
g_s^2 C_g & \simeq g_s^2 C_g(\Lambda) + \frac{\alpha_s}{12\pi} \sum_{q=t,b} \frac{v y^{qq}_S}{m_q} ~,
\\
g_s^2 \tilde C_g & \simeq g_s^2 \tilde C_g(\Lambda) - \frac{\alpha_s}{8\pi} \sum_{q=t,b} \frac{v y^{qq}_P}{m_q} ~,
\end{align}
valid in the limit $m_q \gg m_\phi$,  
in agreement with Higgs low-energy theorems \cite{Shifman:1978zn}.

Among the most stringent constraints on the CP-violating ALP interactions 
stemming from \eq{eq:Lphi}, EDMs emerge as the most powerful probes. 
From the experimental side, there is an outstanding program to improve the current limits on permanent EDMs of molecules, atoms, nuclei and nucleons by several orders of magnitude (see e.g.~\cite{Chupp:2017rkp}). It is therefore mandatory to provide a general framework to systematically account for CP-violating ALP contributions to EDMs, which was missing until recently (see however \cite{Marciano:2016yhf,Stadnik:2017hpa,Dzuba:2018anu}).

The leading low-energy CP-violating Lagrangian relevant for EDMs of molecules, atoms, nuclei and nucleons reads~\cite{Pospelov:2005pr}
\begin{align}
\label{eq:CPV_lagrangian}
\mathcal{L}_{\rm CPV}  & =
\sum_{i,j=u,d,e} \!\!\! C_{ij} (\bar f_i f_i) (\bar f_j i \gamma_5 f_j) 
+ \alpha_s C_{Ge} \, GG \, \bar e i\gamma_5e + \alpha_s C_{\tilde Ge} \, G\tilde G \, \bar e e \\
& 
- \frac{i}{2} \! \sum_{i=u,d,e} d_i \bar f_i (F \!\cdot\! \sigma) \gamma_5 f_i 
- \frac{i}{2} \! \sum_{i=u,d} g_s d^C_i \bar f_i (G \!\cdot\! \sigma) \gamma_5 f_i
+ \frac{d_G}{3} f^{abc} G^a \tilde G^b G^c 
 ~, \nonumber 
\end{align}
where we omitted color-octet 4-quark operators (as they 
emerge only at one-loop level in the ALP framework) and the dim-4 $G\tilde G$ operator. 
The latter is assumed to be absent thanks to a UV mechanism addressing the strong CP problem. 

\begin{figure}[t]
\centering
\includegraphics[width=0.6\linewidth]{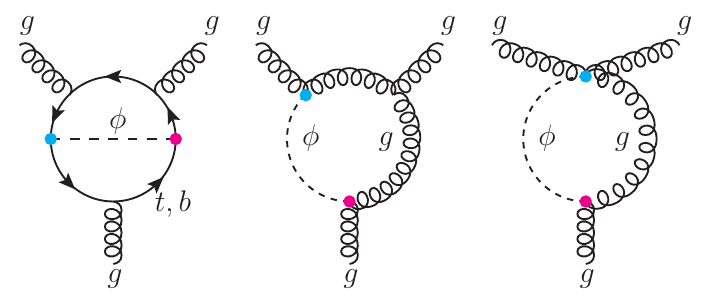}
\caption{Leading contributions to the Weinberg operator. Figure from Ref.~\cite{DiLuzio:2020oah}.}
\label{fig:Weinberg}
\end{figure}

Within our EFT approach, $C_{ij}$, $C_{Ge}$ and $C_{\tilde Ge}$ are generated by the Feynman diagrams 
of Fig.~\ref{fig:4f} and read
\begin{align}
\label{eq:Cij}
C_{ij} \simeq  \frac{v^2}{\Lambda^2} \frac{y^{ii}_S y^{jj}_P}{m^2_\phi}~, \qquad
C_{Ge} = \frac{4\pi}{m^2_\phi} \frac{v}{\Lambda^2} C_g y^{ee}_P ~, \qquad
C_{\tilde Ge} = \frac{4\pi}{m^2_\phi} \frac{v}{\Lambda^2} \tilde C_g y^{ee}_S ~.
\end{align}
The last term of Eq.~(\ref{eq:CPV_lagrangian}) represents the Weinberg operator which is generated by the representative diagrams  
shown in Fig.~\ref{fig:Weinberg}. The related Wilson coefficient $d_G$ reads~\cite{DiLuzio:2020oah}
\begin{align}
d_G & \simeq  \frac{g_s \alpha_s}{(4\pi)^3} \sum_{i=t,b} \! \frac{v^2}{\Lambda^2} \frac{y^{ii}_S y^{ii}_P}{4 m^2_{i}}
+ \frac{3g_s}{\pi^2} \frac{g_s^4 C_g \tilde C_g}{\Lambda^2} \log\frac{\Lambda}{m_\phi} ~ , 
\label{eq:weinberg}
\end{align}
where the first term refers to the two-loop diagram and holds for $m_q \gg m_\phi$.
Instead, the second term of Eq.~(\ref{eq:weinberg}) arises from the one-loop diagrams of Fig.~\ref{fig:Weinberg}.  

\begin{figure}[t]
\centering
\includegraphics[width=0.5\linewidth]{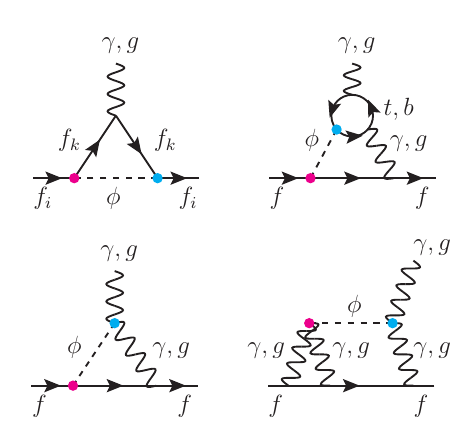}
\caption{Leading contributions to the fermionic (C)EDMs. Figure from Ref.~\cite{DiLuzio:2020oah}.}
\label{fig:dipole}
\end{figure}

Finally, we analyse the fermionic (C)EDMs induced by ALP interactions. The leading contributions stem from the Feynman 
diagrams reported in Fig.~\ref{fig:dipole} and read~\cite{Marciano:2016yhf,DiLuzio:2020oah}
\begin{align}
\frac{d_i}{e} \simeq &
- \sum_{k}
\frac{Q_k}{16 \pi^2}\frac{m_k}{m^2_{\phi}}
\frac{v^2}{\Lambda^2}
\Re (y^{ik}_S y^{ki}_P) \, \ell(x_k)
\nonumber\\
&- \sum_{k} \frac{ N_c \, \alpha \, Q_i \, Q^2_k}{8\pi^3 \, m_k}  
\frac{v^2}{\Lambda^2}
\left( 
y^{ii}_P y^{kk}_S f(x_{k}) + y^{ii}_S y^{kk}_P g(x_{k})  
\right)
\nonumber\\
&-\frac{Q_i}{2\pi^2} \frac{v}{\Lambda^2} e^2 (y^{ii}_S {\tilde C}_\gamma - C_\gamma y^{ii}_P) \log\frac{\Lambda}{m_\phi}
\nonumber\\
&-\frac{3 \, \alpha \, Q^3_i}{\pi^3} \frac{m_i}{\Lambda^2} e^4 C_\gamma {\tilde C}_\gamma \log^2\frac{\Lambda}{m_\phi}
\nonumber\\
& - \delta_{qi} \frac{2 \, \alpha_s \, Q_i}{\pi^3} \frac{m_i}{\Lambda^2} e^2 g^2_s 
( C_\gamma {\tilde C}_g + C_g {\tilde C}_\gamma  )
\log^2\frac{\Lambda}{m_\phi} \, ,
\end{align}
in the EDMs case (where $i=e,u,d$, $q=u,d$, and $N_c =3$) and
\begin{align}
\label{eq:dCi}
d^C_i \simeq &
- \sum_{k}
\frac{1}{16 \pi^2}\frac{m_k}{m^2_{\phi}} \frac{v^2}{\Lambda^2} \Re (y^{ik}_S y^{ki}_P) \, \ell(x_k)
\nonumber\\
&- \sum_{k} \frac{\alpha_s}{16\pi^3 \, m_{k}} \frac{v^2}{\Lambda^2}  
\left( 
y^{ii}_P y^{kk}_S f(x_{k}) + y^{ii}_S y^{kk}_P g(x_{k})  
\right)
\nonumber\\
&-\frac{1}{2\pi^2} \frac{v}{\Lambda^2} g_s^2 (y^{ii}_S {\tilde C}_g - C_g y^{ii}_P) \log\frac{\Lambda}{m_\phi}
\nonumber\\
&-\frac{4 \, \alpha_s}{\pi^3} \frac{m_i}{\Lambda^2} g_s^4 C_g {\tilde C}_g \log^2\frac{\Lambda}{m_\phi}
\nonumber\\
&-\frac{3 \, \alpha \, Q^2_i}{2\pi^3} \frac{m_i}{\Lambda^2} e^2 g_s^2 
( C_\gamma {\tilde C}_g + C_g {\tilde C}_\gamma )
\log^2\frac{\Lambda}{m_\phi} \, ,
\end{align}
for the CEDMs (where $i=u,d$). 
The loop functions are 
$\ell(x) = (3 \!-\! 4x \!+\! x^2 \!+\! 2 \log x) / (1-x)^3$ 
where $x_{k}= m^2_{k} / m^2_\phi$ and, in the limit $x \gg 1$,
$f(x) \approx (6 \log x + 13)/18$ and $g(x) \approx (\log x +2)/2$.
The first diagram of Fig.~\ref{fig:dipole} accounts for 
flavour-violating effects (for flavour-diagonal effects, see~\cite{Chen:2015vqy}) while the second diagram refers 
to Barr-Zee two-loop contributions.

The above expressions have been obtained employing a hard cutoff as a UV regulator and assuming no significant cancellations with finite terms.
Although Eqs.~(\ref{eq:y_l_RGE})-(\ref{eq:dCi}) include only 
leading-order short-distance effects, the bounds of Table~\ref{Tab:ALL} have been derived accounting also for 
one-loop QCD running effects (improved with a two-loop running 
of $\alpha_s$ and quark masses) from $\Lambda = 1$ TeV to $m_\phi= 5\,$GeV and the running of $\mathcal{L}_{\rm CPV}$ from $m_\phi$ to the hadronic scale $\mu_{\rm had} = 1$ GeV~\cite{Degrassi:2005zd}.

Constraints on $d_e$, $C_{ij}$ and $C_{Ge}$ in Eq.~(\ref{eq:CPV_lagrangian}), are obtained by using the polar 
molecule ThO. Indeed, the electron spin-precession frequency 
$\omega_{\text{ThO}}$ is affected by both $d_e$ and CP-odd 
electron--nucleon ($N$) interactions~\cite{Dekens:2018bci}
\begin{align}
\omega_{\text{ThO}} = 1.2 \,\mathrm{mrad}/\mathrm{s}
\left(\frac{d_e}{10^{-29}\,e\,\mathrm{cm}}\right) +1.8 \,\mathrm{mrad}/\mathrm{s}\left(\frac{C_S}{10^{-9}}\right) ~, 
\nonumber
\end{align}
where the theoretical error is within a few percent while the 
experimental limit $\omega_{\text{ThO}} < 1.3\,\mathrm{mrad}/\mathrm{s}$ ($90\%$ C.L.)~\cite{Andreev:2018ayy}.
The coefficient $C_S$ stemming from the interaction $\mathcal{L} \supset - \frac{G_F}{\sqrt{2}} C_S \bar N N \bar e i\gamma_5e$~\cite{Dekens:2018bci} is related to $C_{ij}$ and $C_{Ge}$ as 
$C_S / v^2 \simeq -17 (C_{ue} + C_{de}) + 4.7 \, \text{GeV} \, C_{Ge}$.
The neutron EDM receives contributions from quark (C)EDMs, the Weinberg operator and 4-quark operators~\cite{Hisano:2012cc,Cirigliano:2019vfc,Bertolini:2019out} 
\begin{align}
d_n & = 0.784(28)\,d_u - 0.204(11)\,d_d 
-0.55(28)\,e\, d^C_u  
- 1.10(55)\,e\, d^C_d \notag\\
&+ 50(40)\,{\rm MeV}\,e\,d_{G} + 30(20)\,{\rm MeV} 
\,e\,(C_{ud}-C_{du}) ~ , 
\end{align}
and the current experimental bound is $d_n < 1.8 \cdot 10^{-26}\,e~$cm ($90\%$ C.L.)~\cite{Abel:2020pzs,Pendlebury:2015lrz}. 
Finally, the EDM of the diamagnetic atom ${}^{199}$Hg is generated
by both nuclear and leptonic CP-odd interactions~\cite{Dekens:2018bci,Cirigliano:2019vfc} 
\begin{align}
d_{\rm Hg} \simeq 4.0 \cdot 10^{-4} d_{n} - [ 2.8 \, C_S - 2.1 \, C_P ] \, 10^{-22}\,e\,{\rm cm} ~ , 
\end{align}
with $C_P \simeq C^{(0)}_P - C^{(1)}_P$ defined through the Lagrangian 
$\mathcal{L} \supset - \frac{G_F}{\sqrt{2}} \bar N ( C^{(0)}_P + \tau_3 C^{(1)}_P )  i\gamma_5 N \bar e e$. Our bounds are set by employing
$C_P/v^2 = 350 (C_{eu} + C_{ed}) + 1.1 \, \text{GeV} \, C_{\tilde Ge}$ 
and the experimental limit $d_{\rm Hg}< 6 \cdot 10^{-30}e~$cm ($90\%$ C.L.)~\cite{Graner:2016ses}.

\begin{table}[t]
\small
	\centering
	\begin{tabular}{ccc}
		\toprule
		 CP-violating invariant & Bound & Observable \\
		\colrule
	 	$\abs{C_\gamma {\tilde C}_\gamma}$   & $6.2 \times 10^{-3}$ & 
		$\omega_{\rm ThO}(d_e)$   \\
		$\abs{C_g {\tilde C}_g}$   &  $1.4 \times 10^{-6}$ & $d_n(d_G)$ \\
		$\abs{C_\gamma {\tilde C}_g}$ &  $0.40$  & $d_{\rm Hg}(C_P, C_S)$ \\ 
		$\abs{C_g \tilde C_\gamma}$   & $2.3 \times 10^{-3}$ & $\omega_{\rm ThO}(C_S)$ \\ 
		$\abs{y^{ee}_S {\tilde C}_\gamma - y^{ee}_P C_\gamma}$   &  $6.9 \times 10^{-11}$  &  $\omega_{\rm ThO}(d_e)$  \\	
		$\abs{y^{uu}_S {\tilde C}_g - y^{uu}_P C_g}$ &  $8.1 \times 10^{-9}$  &  $d_n (d^C_{u})$ \\
		$\abs{y^{dd}_S {\tilde C}_g - y^{dd}_P C_g}$ &  $6.5 \times 10^{-9}$  &  $d_n (d^C_{d})$ \\
		$\abs{y^{ee}_P C_g}$   &  $2.1 \times 10^{-11}$  & $\omega_{\rm ThO}(C_S)$ \\ 
		$\abs{y^{ee}_S \tilde C_g}$   &  $7.3 \times 10^{-9}$   & $d_{\rm Hg}(C_P)$ \\ 
		$\abs{y^{uu}_S y^{dd}_P - y^{dd}_S y^{uu}_P}$ &  $5.6 \times 10^{-9}$  &  $d_{\rm Hg} (C_{ud} - C_{du})$ \\
		$\abs{y^{ee}_S y^{uu}_P}$, $\abs{y^{ee}_S y^{dd}_P}$  &  $4.2 \times 10^{-13}$  & $d_{\rm Hg}(C_P)$ \\
		$\abs{y^{uu}_S y^{ee}_P}$, $\abs{y^{dd}_S y^{ee}_P}$   &  $2.1 \times 10^{-13}$ &  $\omega_{\rm ThO}(C_S)$ \\		
		$\abs{y^{tt}_S y^{ee}_P}$   &  $6.8 \times 10^{-9}$  & $\omega_{\rm ThO}(C_S)$ \\
		$\abs{y^{bb}_S y^{ee}_P}$   &  $1.7 \times 10^{-10}$  & $\omega_{\rm ThO}(C_S)$ \\
		$\abs{y^{ee}_S y^{tt}_P}$   &  $8.8 \times 10^{-9}$  &   $\omega_{\rm ThO}(d_e)$ \\
		$\abs{y^{ee}_S y^{bb}_P}$   &  $3.9 \times 10^{-9}$  & $d_{\rm Hg}(C_P)$ \\
		$\abs{y^{tt}_S y^{tt}_P}$   &  $0.10$ &  $d_n(d_G)$  \\
		$\abs{y^{bb}_S y^{bb}_P}$   &  $5.9\times 10^{-5}$ &  $d_n(d_G)$   \\
		$\abs{y^{ee}_S \, y^{ee}_P }$   &  $1.0 \times 10^{-10}$  &  $\omega_{\rm ThO}(d_e)$  \\	
		$\abs{y^{e\mu}_S \, y^{\mu e}_P }$   &  $2.2 \times 10^{-12}$   &  $\omega_{\rm ThO}(d_e)$  \\	
		$\abs{y^{e\tau}_S \, y^{\tau e}_P }$   &  $3.1 \times 10^{-12}$  &  $\omega_{\rm ThO}(d_e)$  \\	
		$\abs{y^{uu}_S \, y^{uu}_P }$   &  $3.9 \times 10^{-8}$  &  $d_n (d^C_{u})$  \\			
		$\abs{y^{uc}_S \, y^{cu}_P }$   &  $3.2 \times 10^{-9}$  &  $d_n (d^C_{u})$  \\	
		$\abs{y^{ut}_S \, y^{tu}_P }$   &  $3.2 \times 10^{-7}$ &  $d_n (d^C_{u})$  \\
		$\abs{y^{dd}_S \, y^{dd}_P }$   &  $4.6 \times 10^{-8}$    &  $d_n (d^C_{d})$  \\			
		$\abs{y^{ds}_S \, y^{sd}_P }$   &  $6.4 \times 10^{-9}$  &  $d_n(d^C_{d})$  \\
		$\abs{y^{db}_S \, y^{bd}_P }$   &  $1.7 \times 10^{-8}$   &  $d_n(d^C_{d})$  \\				
		\botrule
	\end{tabular}\\
	\caption{Bounds on CP-violating invariants for $\Lambda = 1$ TeV and $m_\phi = 5$ GeV. 
	 In the 3rd column we specify the observable and the leading operator setting the bound (in brackets).}
	\label{Tab:ALL}
\end{table}

The EDM's sensitivity to the CP-violating invariants defined in Eq.~(\ref{eq:Jarlskog}) are reported in Table~\ref{Tab:ALL} for $m_\phi = 5$ GeV. 
As we can see, 4-fermion operators set the strongest bounds on several CP invariants. Moreover, the Weinberg operator is particularly effective in constraining ALP couplings 
to top and bottom quarks as well as to gluons. Interestingly, EDMs put severe bounds
also to flavor-violating sources of CP violation that are competitive with those 
stemming from flavour-changing neutral current processes.
Even if the relative impact of the above contributions on the EDMs will depend on the specific ALP model, it is instructive to consider a realistic example where $y^{ii}_{S,P} \propto \frac{m_i}{v}$, $C_{\gamma(g)}$ and $\tilde C_{\gamma(g)} \propto \frac{1}{16\pi^2}$. In this case, 4-fermion operators are the best probes of CP violation followed by the electron EDM and the Weinberg operator which show similar sensitivities. 
Let us also remark that the electron EDM bounds are always stronger than the limits from the anomalous magnetic moment of the electron, unless CP-violating phases are smaller than about $10^{-4}$~\cite{Giudice:2012ms}.
Future experimental projections are particularly promising. 
The experimental bound on the neutron EDM measurement should reach the level 
of $d_n \lesssim 10^{-28}e~$cm~\cite{Chupp:2017rkp}. There are also plans 
to measure the proton and deuteron EDMs in electromagnetic storage rings at 
the level of $d_{p,D} \lesssim 10^{-29}e~$cm~\cite{Chupp:2017rkp}. 
Furthermore, we expect also one order of magnitude improvement on the current measurement of molecular systems, e.g. the polar molecule ThO, which currently sets the most powerful bounds on the electron EDM and electron-nucleon couplings. 
If the above forecasts are respected, the limits on $d_G$ and quark (C)EDMs will 
improve by roughly three orders of magnitude while the constraints on electron EDM 
and 4-fermion contributions will be strengthened by one order of magnitude.

\subsection{CP-violating ALP Lagrangian: $m_\phi \lesssim 1$ GeV} 
\label{sec:ALPbelow1GeV}

In this section, we are going to extend the previous analysis to ALP masses in the sub-GeV region, 
where non-perturbative methods, such as chiral perturbation theory ($\chi$pt), have to be employed. 
The construction of the effective chiral Lagrangian for an ALP interacting with photons and light pseudoscalar mesons has been 
already discussed at length in the literature for the case of a 
CP-odd ALP (see Refs.~\cite{Georgi:1986df,Bauer:2020jbp,Bauer:2021wjo,Bandyopadhyay:2021wbb}, 
as well as \cite{GrillidiCortona:2015jxo,Vonk:2020zfh,Vonk:2021sit,DiLuzio:2021vjd,DiLuzio:2022tbb} 
for analyses beyond leading order in $\chi \text{pt}$).  
The development of the chiral Lagrangian for 
CP-even scalar interactions was elaborated instead in Ref.~\cite{Leutwyler:1989tn}. 
While we refer to Ref.~\cite{DiLuzio:2023edk} for the construction of the 
most general chiral Lagrangian at $\mathcal{O}(p^{2})$ describing 
CP-violating interactions of a light ALP with mesons and baryons, we report here the main results needed 
to estimate the ALP contribution 
to the EDMs. 

To match the notation of Ref.~\cite{DiLuzio:2023edk} we start from a slightly different version of the ALP effective Lagrangian, 
compared to the one in \eq{eq:Lphi}, featuring a derivative ALP coupling to the quark axial current, namely  
\begin{align}
\label{eq:AAA}
    \mathcal{L}_\phi^{\text{QCD}} &= 
    e^2 \frac{\tilde{C}_\gamma}{\Lambda} \phi F \tilde F
    + g_s^2 \frac{\tilde{C}_g}{\Lambda} \phi G \tilde G
    + \frac{\partial_\mu \phi}{\Lambda} \bar{q} \, \gamma^\mu Y_P \gamma_5 \, q
    \nonumber \\
    &+ e^2 \frac{C_\gamma}{\Lambda} \phi F F
    + g_s^2 \frac{C_g}{\Lambda} \phi G G 
 + \frac{v}{\Lambda} \, \phi \,  \bar{q} \, y_{S}\,  q  
 ~ ,
\end{align}
that is valid above the GeV scale. 
Here, $q^T = (u,d)$, while $Y_P$ 
and $y_{S}$ are hermitian matrices.  

Upon taking into account hadronization effects, following a chiral representation of the ALP interactions 
with quarks and gluons, 
Ref.~\cite{DiLuzio:2023edk} obtained a low-energy effective Lagrangian describing the interaction of the ALP with 
light particles: photons, leptons, pions and nucleons. 
The ALP effective Lagrangian can be cast in two contributions displaying opposite CP transformation properties, given respectively by (in the case of two active flavours, $N_f=2$)~\cite{DiLuzio:2023edk}
\begin{align}
\label{eq:Simplest_Lag_2f_CPE}
\mathcal{L}_{\phi, \text{\,CP-even}}^{\chi\text{pt}} &= e^2 \frac{c_\gamma}{\Lambda} \, \phi \,  FF + \frac{v}{\Lambda} y_{\ell, S}^{ij}\,  \phi \, \bar{\ell}_i \ell_j +  \kappa \, \frac{\phi}{\Lambda} \, \left[\partial^\mu \pi_0 \partial_\mu \pi_0 + 2 \partial^\mu \pi^+ \partial_\mu \pi^-\right] \nonumber \\
&+ \frac{\phi}{\Lambda}\,  (\pi_0^2 + 2\pi^+ \pi^-) m_\pi^2 \left[\frac{v}{2} \frac{\mathcal{Z}_u + \mathcal{Z}_d}{m_u + m_d} - 2 \, \kappa \right]  + \frac{\phi}{\Lambda} C_{\phi \text{NN}} \bar{N} N
\end{align}
and
\begin{align}
\label{eq:Simplest_Lag_2f_CPO}
\mathcal{L}_{\phi, \text{\,CP-odd}}^{\chi\text{pt}} &=  e^2 \frac{\tilde{c}_\gamma}{\Lambda} \, \phi \, F\tilde{F}+ i \frac{v}{\Lambda} y_{\ell, P}^{ij}\,  \phi \, \bar{\ell}_i \gamma_5 \ell_j + \frac{\phi}{\Lambda}\, \pi_0 (\pi_0^2 + 2\pi^+ \pi^-)  \left[ \frac{m_\pi^2 m_\phi^2}{m_\pi^2-m_\phi^2}\frac{\Delta^A_{ud}}{6 f_\pi \Lambda} \right] \nonumber \\
&+ \frac{2}{3}\frac{\Delta^A_{ud}}{f_\pi \Lambda} \frac{m_\pi^2}{m_\pi^2- m_\phi^2} \, \partial^\mu \phi\,   (\pi_0 \pi^- \partial_\mu \pi^+ + \pi_0 \pi^+ \partial_\mu \pi^- - 2\pi^+ \pi^- \partial_\mu \pi^0) \nonumber \\
& + \frac{\partial_\mu \phi}{\Lambda} \bar{N} \tilde{C}_{\phi\text{N}}\gamma^\mu \gamma_5 N ~ ,   
\end{align}
where explicit expressions for the parameters appearing in the above Lagrangians 
in terms of the short-distance parameters of \eq{eq:AAA}  
are reported below in Eq. \eqref{eq:BBB}. 
In particular, the physical ALP mass after diagonalization of the ALP-pion system reads~\cite{DiLuzio:2023edk} 
\begin{equation}
m^2_{\phi} = M_\phi^2 + m_\pi^2 (\lambda_g)^2\frac{m_um_d}{(m_u + m_d)^2} \frac{f_\pi^2}{\Lambda^2} + m_\pi^2 \frac{M_\phi^4 }{(m_\pi^2-M_\phi^2)^2}(\Delta^A_{ud})^2\frac{f_\pi^2}{\Lambda^2} ~ , 
\end{equation}
where $M_\phi$ is the bare ALP mass and $\lambda_g = 32 \pi^2 \tilde{C}_g$.

\begin{table}[ht!]
    \centering
    \begin{tabular}{|c|c|c|c|c|c|}
    \hline
                            & $c_\gamma$ & $y_{\ell,S}$&  $\kappa$& $\mathcal{Z}$& $C_{\phi \text{NN}}$ \\
                            \hline
      $\tilde{c}_\gamma$    & $ \tilde{c}_\gamma\,  c_\gamma $ & $ \tilde{c}_\gamma\,  y_{\ell,S}$ & $ \tilde{c}_\gamma\,  \kappa $ & $ \tilde{c}_\gamma\,  \mathcal{Z} $ & $ \tilde{c}_\gamma\, C_{\phi \text{NN}} $\\
      $y_{\ell, P}$    & $ y_{\ell, P}\,  c_\gamma $ & $ y_{\ell, P}\,  y_{\ell,S}$ & $ y_{\ell, P}\,  \kappa $ & $ y_{\ell, P}\,  \mathcal{Z} $ & $ y_{\ell, P}\,  C_{\phi \text{NN}} $\\
      $\Delta_{ud}^A$    & $ \Delta_{ud}^A\,  c_\gamma $ & $ \Delta_{ud}^A\,  y_{\ell,S}$ & $ \Delta_{ud}^A\,  \kappa $ & $ \Delta_{ud}^A\,  \mathcal{Z} $ & $ \Delta_{ud}^A\, C_{\phi \text{NN}} $\\
      $\tilde{C}_{\phi\text{N}}$    & $ \tilde{C}_{\phi\text{N}}\, c_\gamma $ & $ \tilde{C}_{\phi\text{N}}\, y_{\ell,S}$ & $ \tilde{C}_{\phi\text{N}}\, \kappa $ & $ \tilde{C}_{\phi\text{N}}\,\mathcal{Z} $ & $ \tilde{C}_{\phi\text{N}}\, C_{\phi \text{NN}} $\\
      \hline
    \end{tabular}
    \caption{Jarlskog invariants emerging from the interactions in Eq.~\eqref{eq:Simplest_Lag_2f_CPE} 
    and \eqref{eq:Simplest_Lag_2f_CPO}. 
}
    \label{tab:2F_Jarlskog}
\end{table}
Note that, independently of the scalar or pseudo-scalar nature of the ALP field, CP violation is unavoidable as long as at least one CP-even and one CP-odd couplings from \eq{eq:Simplest_Lag_2f_CPE} and (\ref{eq:Simplest_Lag_2f_CPO}) are simultaneously 
non-vanishing.
CP-violating effects can be described in terms of 
low-energy
Jarlsokg invariants, which are reported in Table \ref{tab:2F_Jarlskog}.
In particular, the coefficients 
appearing in Table \ref{tab:2F_Jarlskog} 
are related to the microscopic parameters of Eq.~(\ref{eq:AAA}) via the following relations~\cite{DiLuzio:2023edk}
\begin{align}
\label{eq:BBB}
    c_\gamma  &= C_\gamma - \frac{\beta^0_{\text{QED}}}{\beta^0_{\text{QCD}}}  C_g ~ ,\qquad
    \tilde{c}_\gamma = \, \tilde{C}_\gamma  -4 N_c \text{ tr }(Q_AQ_q^2)  \tilde{C}_g ~ , \qquad 
    \kappa = \frac{8\pi}{\alpha_s}\frac{g_s^2 C_g}{\beta^0_\text{QCD}} 
     ~ , \nonumber \\ 
    \mathcal{Z} &= y_{S}  + \frac{g_s^2 C_g }{\beta^0_{\text{QCD}}} \frac{8\pi}{\alpha_s} \frac{M_q}{v} ~ , \qquad 
    \Delta^A_{ud} = Y_P^u - Y_P^d -  16 \pi^2 \tilde{C}_g \frac{m_d-m_u}{m_d+m_u} ~, \nonumber \\
    \tilde{C}_{\phi\text{p}} &=  Y_P^u \Delta_u +Y_P^d \Delta_d -  16 \pi^2 \tilde{C}_g  \left[ \frac{m_u \Delta_d}{m_u + m_d} + \frac{m_d \Delta_u}{m_u + m_d} - \frac{m_\phi^2 \Delta^A_{ud}}{m_\pi^2-m_\phi^2} \frac{\Delta_u-\Delta_d}{32 \pi^2 \tilde{C}_g } \right] 
    ~ , \nonumber \\
    \tilde{C}_{\phi\text{n}} &=  Y_P^u \Delta_d - Y_P^d \Delta_u -  16 \pi^2 \tilde{C}_g \left[ \frac{m_u \Delta_u}{m_u + m_d} + \frac{m_d \Delta_d}{m_u + m_d} + \frac{m_\phi^2 \Delta^A_{ud}}{m_\pi^2-m_\phi^2} \frac{\Delta_u-\Delta_d}{ 32 \pi^2 \tilde{C}_g } \right]
    ~ , \nonumber \\
    C_{\phi \text{pp}} &=  \frac{y_{q,S}^u \, v}{m_u}  \, \sigma_u +  \frac{y_{q,S}^d \, v}{m_d} \, \sigma_d +  \frac{y_{q,S}^s \, v}{m_s} \,  \sigma_s - \frac{32\pi^2 C_g}{9} \,  (m_p - \sigma_u -\sigma_d-\sigma_s) ~ ,  \nonumber \\
    C_{\phi \text{nn}} &=  \frac{y_{q,S}^u \, v}{m_d} \, \sigma_d +  \frac{y_{q,S}^d \, v}{m_u} \, \sigma_u +  \frac{y_{q,S}^s \, v}{m_s} \,  \sigma_s - \frac{32\pi^2 C_g}{9} \,  (m_p - \sigma_u -\sigma_d-\sigma_s) ~ , 
\end{align}
with the low-energy QCD matrix elements given by $\Delta_u = 0.858(22)$ and $\Delta_d = -0.418(22)$ \cite{DiLuzio:2023tqe,ParticleDataGroup:2022pth}, 
while the values of 
$\sigma_u$, 
$\sigma_d$ and $\sigma_s$ can be found e.g.~in Ref.~\cite{Cheng:2012qr}. 

\subsubsection{ALP contribution to EDMs} 
\begin{figure}[t]
\centering
\includegraphics[width=1.0\linewidth]{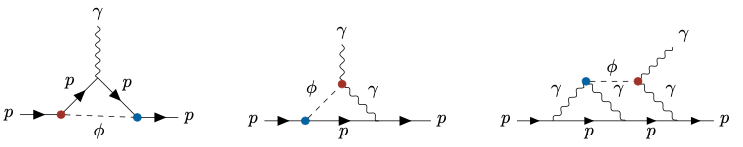}
\caption{Feynman diagrams contributing to the ALP-mediated effects to the proton EDM, $d_p$. 
Figure from Ref.~\cite{DiLuzio:2023edk}.}
\label{fig:pEDMALPDiagrams}
\end{figure}

The effective Lagrangian provided in \eqs{eq:Simplest_Lag_2f_CPE}{eq:Simplest_Lag_2f_CPO} 
readily allows one to compute the contributions of a 
light ALP to the proton EDM, with   
the relevant topologies displayed in Fig.~\ref{fig:pEDMALPDiagrams}.
Following the standard definition of the EDM of a fermion given in \eq{eq:CPV_lagrangian},  
one finds~\cite{DiLuzio:2023edk}
\begin{align}
d_p &\simeq -\frac{e \, Q_p}{4\pi^2 \Lambda^2} 
\left[C_{\phi\text{pp}}\tilde{C}_{\phi \text{p}}  + 
e^2 m_p c_\gamma \tilde{C}_{\phi \text{p}} 
\left( 6 + 2\ln \frac{\Lambda_{\text{ren}}^2}{m_p^2}\right) +   e^2 \tilde{c}_\gamma C_{\phi \text{pp}} 
\left(2 + \ln \frac{\Lambda_{\text{ren}}^2}{m_p^2} \right) \right. \nonumber \\
&\left. + 3 \frac{Q_p^2}{\pi^2} m_p e^6 c_\gamma \tilde{c}_\gamma \ln^2 \frac{\Lambda_{\text{ren}}}{m_\phi}
\right] ~ ,
\end{align}
in the limit $m_\phi \ll m_p$ and setting the renormalization scale $\Lambda_{\rm ren}\simeq m_p$. 

Due to their lack of a minimal coupling to the electromagnetic field-strength tensor, neutrons cannot develop an EDM at 
leading order in $\chi$pt. However, they do display non-minimal couplings to photons at next-to-leading order in $\chi$pt \cite{Fettes:1998ud,Oller:2006yh} which can eventually induce a non-zero value for $d_n$.  
In the $N_f=2$ case, the 
relevant relativistic Lagrangian reads
\begin{equation}
\mathcal{L}_{\gamma \text{N}}^{\text{NLO}} = 
\frac{F_{\mu\nu}}{4m_N}\left[\, \mathcal{C}_p \,\bar{p}\sigma^{\mu\nu}p + \mathcal{C}_n \,\bar{n}\sigma^{\mu\nu}n \,\right] ~ , 
\end{equation} 
where $\mathcal{C}_p \simeq 1.79$~\cite{ParticleDataGroup:2022pth} and $\mathcal{C}_n \simeq 1.91$~\cite{ParticleDataGroup:2022pth} are measured low-energy constants.
Given these interactions, one can then compute the corresponding contribution to $d_n$, whose Feynman 
diagrams are the same as for the proton
case, finding the following order of magnitude estimate~\cite{DiLuzio:2023edk}
\begin{equation}
d_n \sim -\frac{e \, \mathcal{C}_n}{4\pi^2 \Lambda^2} 
\left[ -\frac{3}{8}C_{\phi\text{nn}}\tilde{C}_{\phi \text{n}}
\frac{\Lambda_{\text{ren}}^2}{m_n^2} 
+ e^2 (2 m_n c_\gamma \tilde{C}_{\phi \text{n}}+ 3 \tilde{c}_\gamma C_{\phi \text{nn}}) \ln \frac{\Lambda_{\text{ren}}^2}{m_n^2} \right]~.
\end{equation}
Note that $d_p$ and $d_n$ have a comparable size while their
current experimental bounds read $d_p < 2.1 \times 10^{-25}\,e$ cm~\cite{Sahoo:2016zvr}
and $d_n < 1.8 \times 10^{-26}\,e$ cm (90\% C.L.)~\cite{Abel:2020pzs,Pendlebury:2015lrz}.

Constraints on the low-energy Jarlskog invariants are obtained by imposing the existing bounds on $d_p$ and $d_n$.
Turning on just two couplings at a time, one finds the  
limits displayed in Table \ref{tab:constraints}.
\begin{table}[ht!]
    \centering
    \begin{tabular}{|c|c|c|c|c|}
    \hline
                           & $y_S^u$ & $y_S^d$ & $C_\gamma$ & $C_g$\\
                           \hline
      $y_P^u$                & $ 2.8 \times 10^{-12} $ & $ 1.8 \times 10^{-12}$   & $ 8.0 \times 10^{-8}$  & $ 1.1 \times 10^{-10}$\\
      \hline
      $y_P^d$                & $ 9.9 \times 10^{-12}$ & $ 6.4 \times 10^{-12} $  & $ 1.1 \times 10^{-7}$  & $ 3.8 \times 10^{-10}$\\
      \hline
      $\tilde{C}_\gamma$     & $ 1.2 \times 10^{-6}$ & $ 1.8 \times 10^{-6}$  & $  4.8\times 10^{-3}$ & $ 7.1 \times 10^{-5}$\\
      \hline
      $\tilde{C}_g$          & $ 1.1 \times 10^{-9}$ & $ 7.4 \times 10^{-10}$ & $ 2.4 \times 10^{-5}$  & $ 4.4 \times 10^{-8}$\\ \hline
    \end{tabular}
    \caption{Upper limits on Jarlskog invariants obtained 
    from the bounds on the neutron and proton EDMs assuming $\Lambda = 1$ TeV~\cite{DiLuzio:2023edk}. To make contact with the notation in \eq{eq:Lphi} here we have defined $y_{P}^{q} = - 2\frac{m_q}{v} Y_P^q$. 
    }
    \label{tab:constraints}
\end{table}

\subsection{Towards a UV completion for the CP-violating ALP} 
\label{sec:UVALP}

It is interesting to speculate on the possible origin of the 
CP-violating ALP effective Lagrangian in \eq{eq:Lphi}. 
The underlying UV dynamics can be naturally conceived in the framework of 
strongly-coupled theories. A paradigmatic example is given 
by the SM itself, when considering the effective interactions of the neutral pion $\pi^0$ below the GeV scale 
(see e.g.~\cite{Choi:1990cn}). In this setup, the role of the ALP is played by $\pi^0$, quark masses induce the breaking of the shift symmetry while the 
QCD $\theta$ term provides the necessary CP-violating source. Finally, electromagnetic interactions
act as mediators from the strong sector to the $\pi^0$. CP-odd pion interactions 
include the terms 
\beq 
A_1 \frac{\pi^0}{f_\pi} F \tilde F + A_2 \frac{\partial_\mu \pi^0}{f_\pi} \bar e \gamma^\mu \gamma_5 e ~, \\
\eeq 
where $A_1 = \frac{\alpha}{4\pi}$ is the Wess-Zumino-Witten term and $A_2$ is radiatively induced by $A_1$ via electromagnetic interactions, so that 
$A_2 \sim (\frac{\alpha}{4\pi})^2$. CP-even pion interactions
\beq 
C_1 \frac{\pi^0}{f_\pi} F F + C_2 m_e \frac{\pi^0}{f_\pi} \bar e e ~ , 
\eeq
are instead sourced by the QCD $\theta$ term. 
In particular, $C_1$ emerges from integrating out heavy charged-baryon loops. 
The CP-even Yukawa-like couplings between pions and baryons appearing in such triangle loops naturally emerge from the baryonic Lagrangian upon including the $\theta$ term in the phase of the quark mass matrix. 
$C_2$ is in turn generated radiatively from $C_1$ via electromagnetic
interactions and thus scales as $C_2 \sim \theta \frac{\alpha}{4\pi}$.
Strongly coupled new physics scenarios emerging at the 
scale $\Lambda \gtrsim 1$ TeV, that mimic the pion dynamics of the SM, 
can hence be conceived in analogy.

Other interesting scenarios motivating the study of a CP-violating ALP are provided by relaxion models \cite{Graham:2015cka}, aiming at solving the hierarchy problem through an ALP field, the relaxion, which scans the Higgs boson mass in the early Universe from a very high-energy scale $\mu \gg 1$ TeV down to the final value of $\mathcal{O}(v)$. 
The presence of both the relaxion-Higgs mixing and the relaxion-photon/gluon couplings does violate CP.
As an explicit example let us consider the following relaxion Lagrangian 
\begin{align}
\label{eq:ALP_Lag_Above_EW_v1}
\mathcal{L}_{\text{Relax}} &= \frac{1}{2} \partial^\mu \Phi \partial_\mu \Phi  - \frac{\Phi}{\Lambda} \left(\tilde{\mathcal{C}}_g\, g_s^2\, G \tilde{G} + \tilde{\mathcal{C}}_W\,g^2\, W \tilde{W} + \tilde{\mathcal{C}}_B\,g'^2\, B \tilde{B} \right)\nonumber \\
&+ \frac{\partial^\mu \Phi}{\Lambda} \sum_f \bar{\psi}_f \gamma_\mu \tilde{X}_f \psi_f- V(H, \Phi) ~ , 
\end{align}
where $\Phi$ is the relaxion field and $f = \left\{ \ell_L, q_L, u_R, d_R, e_R\right\}$ are the chiral SM fermions,  
and the relaxion-Higgs potential is given by 
(see e.g.~\cite{Graham:2015cka,Espinosa:2015eda,Choi:2016luu,Flacke:2016szy}) 
\begin{align}
\label{eq:Higgs_ALP_Potential}
    V (H, \Phi ) &= \left[ - \mu^2 + \tilde{g} \mu \Phi \right] |H^\dagger H| - M^2 |H^\dagger H| \cos \left(\frac{\Phi}{\Lambda}\right) \nonumber \\
    & + \lambda \, |H^\dagger H|^2  + r \tilde{g} \mu^3 \Phi + (m_\phi^0)^2 \Phi^2 + \dots ~ .
\end{align}
Of these terms, the linear one in $\Phi$ 
prompts the cosmological evolution of the relaxion field, which allows for scanning of the Higgs VEV via the relaxion-dependent Higgs mass parameter. The cosine term 
acts instead as a backreaction, producing the potential barrier that eventually halts the evolution of the relaxion field around 
the Higgs VEV. 
Expanding both the Higgs and relaxion fields around their VEV,
\begin{equation}
 \Phi = \phi_0 + \phi' \qquad \text{and} \qquad H = \frac{1}{\sqrt{2}}\begin{pmatrix} 0\\ v + h'\end{pmatrix} ~ , 
\end{equation}
generates in turn a mixing between the Higgs and the relaxion fields, with the mass eigenstates, $\phi = c_\theta \phi' + s_\theta h' $ and $h = c_\theta h' - s_\theta \phi'$, 
given in terms of the mixing angle $s_\theta \simeq \frac{1}{2\lambda}\frac{M^2}{v^2}\frac{v}{\Lambda} \sin \frac{\phi_0}{\Lambda} \ll 1$. The resulting interaction 
Lagrangian in the mass eigenbasis at $\mathcal{O}(1/\Lambda)$  is given by  
\begin{align}
\label{eq:ALP_Lagrangian_In_The_Broken_Phase_At_EWSB}
     \mathcal{L}_{\text{Relax}} &\supset  c_\theta \frac{\partial_\mu \phi}{\Lambda} \sum_f \bar{f} \,\gamma^\mu (X_f^V + X_f^A \gamma_5) f  - s_\theta \frac{\phi}{v} \sum_f \frac{m_f}{v}\bar{f} f  \nonumber \\
     &   -c_\theta\frac{\phi}{\Lambda} \Bigg[\tilde{\mathcal{C}}_g\,g_s^2\, G \tilde{G}  + e^2 \Big( \tilde{\mathcal{C}}_\gamma F \tilde{F} + 2 \frac{\tilde{\mathcal{C}}_{\gamma Z}}{s_wc_w}F \tilde{Z}  + \frac{\tilde{\mathcal{C}}_Z}{s_w^2 c_w^2} Z \tilde{Z} \Big) \nonumber\\
     & +\,  \tilde{\mathcal{C}}_W \,g^2\,\left( W^{+\mu\nu}\tilde{W}^-_{\mu\nu} - 2 i g (s_w A_\nu + c_w Z_\nu) W_\mu^+ \tilde{W}^{-\mu\nu} \right.\nonumber\\
     & \left.+ 2 i g  W_\mu^+W_\nu^-(c_w \tilde{F}^{\mu\nu} + s_w \tilde{Z}^{\mu\nu}) + \text{h.c.}\right) \Bigg] + s_\theta \frac{\phi}{v} \left(\frac{2}{3}\mathcal{C}_g \,g_s^2\, G G+ 2 \mathcal{C}_\gamma e^2 FF\right) \nonumber \\
     &  - \left(\varphi^+ \varphi^- + \frac{\varphi_Z^2}{2}\right) \left[ c_\theta\frac{\phi}{\Lambda} M^2 \sin \frac{\phi_0}{\Lambda}  + 2 s_\theta \lambda \phi v + 2 c_\theta s_\theta \phi h\right]  \nonumber\\
     & + 2 s_\theta \frac{\phi}{v} \left(\!m_W^2 W^+_\mu W^{-\mu} \!+ \frac{m_Z^2}{2} Z_\mu Z^\mu \right) \!+ 2 s_\theta c_\theta \frac{\phi h }{v^2} \left(\! m_W^2 W_\mu^+ W^{-\mu}\! + \frac{m_Z^2}{2} Z_\mu Z^\mu\right) \nonumber \\
     & - 3 \lambda c_\theta^2 s_\theta h^2 \phi - \lambda c_\theta^3 s_\theta h^3 \phi- c_\theta^3 \frac{\phi h^2}{\Lambda} M^2 \sin \frac{\phi_0}{\Lambda} + \mathcal{L}_{\text{Higgs}}^{\text{kin}} (c_\theta h \rightarrow s_\theta \phi) ~ ,
\end{align}
where $ \mathcal{L}_{\text{Higgs}}^{\text{kin}}$ is the Lagrangian piece containing all the interactions between the Higgs boson, the electroweak gauge bosons and the respective Goldstones that stem from the Higgs kinetic term.  
Moreover, we have defined $2X_{V/A}^f = (K_f^\dagger \tilde{X}_R^f K_f \, \pm \, U_f^\dagger \tilde{X}_L^f U_f) $, where $K_f$ and $U_f$ are the right-handed and left-handed
fermion rotation matrices needed to go from the interaction basis to the mass basis.

The Lagrangian in \eq{eq:ALP_Lagrangian_In_The_Broken_Phase_At_EWSB} has then to be evolved down to the QCD scale. This procedure includes both
integrating out heavy particle states and the 
running of Wilson coefficients \cite{DasBakshi:2023lca}. 
Finally, the resulting Lagrangian can be matched onto the effective one in \eq{eq:AAA}, yielding 
\begin{align}
 C_\gamma &= 2 \,s_\theta\, \mathcal{C}_\gamma \frac{\Lambda}{v} ~ , \qquad 
 C_g = \frac{2}{3} \,s_\theta\, \mathcal{C}_g \frac{\Lambda}{v} ~ , \\ 
\tilde{C}_\gamma &= - 2 \,c_\theta\, \tilde{\mathcal{C}}_\gamma ~ , \qquad \, 
\tilde{C}_g = -\,c_\theta\, \tilde{\mathcal{C}}_g ~ , \\ 
y_s &= -s_\theta\, \frac{M_q}{v}\frac{\Lambda}{v} ~ , \quad \,
Y_P = c_\theta\,\frac{K_f^\dagger \tilde{X}_R^f K_f-U_f^\dagger \tilde{X}_L^f U_f}{2} ~ .
\end{align}

\section{Conclusions}
\label{sec:concl}

In this review, we explored the theoretical landscape 
of CP-violating axions, delving into both the well-established framework of QCD axions and the broader setup of ALPs. Our discussion centred around two main themes: the implications of CP violation for axion-mediated forces 
in the context of QCD axions 
and the effective interactions of CP-violating ALPs, emphasizing their contributions to 
EDMs.

The structural sensitivity of the QCD axion 
to high-energy sources of CP violation, as shown by the  estimate in \eq{eq:thetaeffmin3}, 
suggests rethinking the role of the axion, 
to be considered not just as a ``laundry soap'' of CP violation in strong interactions, 
but rather as a low-energy portal to high-energy sources of CP violation,   
thus turning the 
``strong CP problem'' into the ``strong CP opportunity''.
In particular, 
axion-mediated forces, 
which are enhanced by the presence of a 
CP-violating axion coupling to nucleons, 
provide an extra experimental handle 
for probing UV sources of CP violation in the quark sector, 
with projected sensitivities that are stronger 
(in a given range of axion masses, cf.~\fig{fig:mondip})
than current EDM limits. 
Moreover, if the origin of CP-violating axion couplings
does not stem from QCD-coupled physics, like e.g.~in the 
presence of PQ-breaking operators, the CP-violating axion potential is unsuppressed above the temperatures 
of the QCD crossover,  
and the relative enhancement compared to the 
CP-preserving QCD potential might lead to sizeable deviations 
in the axion relic abundance from the misalignment mechanism. However, as discussed in \sect{sec:CPVaxioncosmo}, the relative enhancement is only sufficient to achieve such deviation in the regime in which the axion provides a sub-fraction of dark matter, unless tuning is invoked to relax the nEDM constraint. 

In the second part, we reviewed the contribution of CP-violating ALP interactions to EDMs of nucleons, nuclei, atoms and molecules, following two recent works \cite{DiLuzio:2020oah,DiLuzio:2023edk}, respectively for the case of ALP masses above and below the GeV scale. In the latter case, 
we employed a chiral Lagrangian approach to parametrize the general CP-violating interactions of the ALP with pions and nucleons. 

The analysis of CP-violating ALPs could be expanded in several directions. For instance, it could be worth to construct explicit UV completions of the CP-violating ALP effective Lagrangian, which were briefly sketched in \sect{sec:UVALP}.  
Another interesting path would be to investigate whether a successful baryogenesis mechanism can be 
driven by an ALP field \cite{Jeong:2018ucz,Jeong:2018jqe,Domcke:2020kcp,Im:2021xoy,Harigaya:2023bmp}, thus providing a 
rationale for exploring CP-violating ALP interactions.

\section*{Acknowledgments}
The work of LDL and PP 
is supported
by the European Union -- Next Generation EU and
by the Italian Ministry of University and Research (MUR) 
via the PRIN 2022 project n.~2022K4B58X -- AxionOrigins. 
The work of LDL, HG and PS is funded by the European Union -- Next Generation EU and by the University of Padua under the 2021 STARS Grants@Unipd programme (Acronym and title of the project: CPV-Axion -- Discovering the CP-violating axion).


\bibliographystyle{ws-wsarpp}
\bibliography{sample}

\end{document}